\documentclass{aastex63}

\usepackage{amsmath}
\received{June 1, 2019}
\revised{January 10, 2019}
\accepted{\today}
\submitjournal{ApJ}

\begin{document}
\shortauthors{Li et al.}
\shorttitle{Probing the full CO SLED of a quasar-starburst system at $z=6.003$}
\title{Probing the full CO spectral line energy distribution (SLED) in the nuclear region of a quasar-starburst system at $z=6.003$}

\correspondingauthor{Jianan Li}
\email{jiananl@pku.edu.cn}
\correspondingauthor{Ran Wang}
\email{rwangkiaa@pku.edu.cn}

\author{Jianan Li}
\affiliation{Department of Astronomy, School of Physics, Peking University, Beijing 100871, China}
\affiliation{Kavli Institute for Astronomy and Astrophysics, Peking University, Beijing 100871, China}

\author{Ran Wang}
\affiliation{Kavli Institute for Astronomy and Astrophysics, Peking University, Beijing 100871, China}

\author{Dominik Riechers}
\affiliation{Department of Astronomy, Cornell University, Space Sciences Building, Ithaca, NY 14853, USA}
\affiliation{Max-Planck-Institut for Astronomie, K\"{o}nigstuhl 17, D-69117 Heidelberg, Germany}

\author{Fabian Walter}
\affiliation{Max-Planck-Institut for Astronomie, K\"{o}nigstuhl 17, D-69117 Heidelberg, Germany}

\author{Roberto Decarli}
\affiliation{INAF -- Osservatorio di Astrofisica e Scienza dello Spazio, via Gobetti 93/3, 40129 Bologna, Italy}

\author{Bram P. Venamans}
\affiliation{Max-Planck-Institut for Astronomie, K\"{o}nigstuhl 17, D-69117 Heidelberg, Germany}

\author{Roberto Neri}
\affiliation{Institute de Radioastronomie Millimetrique, St. Martin d'Heres, F-38406, France}

\author{Yali Shao}
\affiliation{Department of Astronomy, School of Physics, Peking University, Beijing 100871, China}
\affiliation{Kavli Institute for Astronomy and Astrophysics, Peking University, Beijing 100871, China}

\author{Xiaohui Fan}
\affiliation{Steward Observatory, University of Arizona, 933 North Cherry Avenue, Tucson, AZ 85721, USA}

\author{Yu Gao}
\affiliation{Purple Mountain Observatory $\&$ Key Laboratory for Radio Astronomy, Chinese Academy of Sciences, 10 Yuanhua Road, Nanjing
210033, PR China}

\author{Chris L. carilli}
\affiliation{Cavendish Laboratory, 19 J. J. Thomson Avenue, Cambridge CB3 0HE, UK}
\affiliation{National Radio Astronomy Observatory, Socorro, NM 87801-0387, USA}

\author{Alain Omont}
\affiliation{Institut d'Astrophysique de Paris, Sorbonne Universit\'{e}, CNRS, UMR 7095, 98 bis bd Arago, 75014 Paris, France}

\author{Pierre Cox}
\affiliation{Institut d'Astrophysique de Paris, Sorbonne Universit\'{e}, CNRS, UMR 7095, 98 bis bd Arago, 75014 Paris, France}

\author{Karl M. Menten}
\affiliation{Max-Planck-Institut f{\"u}r Radioastronomie, Auf dem H\"{u}gel 69, 53121 Bonn, Germany}

\author{Jeff Wagg}
\affiliation{SKA Organization, Lower Withington Macclesfield, Cheshire SK11 9DL, UK}

\author{Frank Bertoldi}
\affiliation{Argelander-Institut f{\"u}r Astronomie, University at Bonn, Auf dem H\"{u}gel 71, D-53121 Bonn, Germany}

\author{Desika Narayanan}
\affiliation{Department of Astronomy, University of Florida, 211 Bryant Space Science Center, Gainesville, FL 32611, USA}

\begin{abstract}
We report Atacama Large Millimeter/submillimeter Array (ALMA) observations of CO $(8-7)$, $(9-8)$, $\rm H_{2}O (2_{0,2}-1_{1,1})$ and $\rm OH^{+} (1_{1}-0_{1})$ and NOrthern Extended Millimeter Array (NOEMA) observations of CO $(5-4)$, $(6-5)$, $(12-11)$ and $(13-12)$ towards the $z = 6.003$ quasar SDSS J231038.88+185519.7, aiming to probe the physical conditions of the molecular gas content of this source. 
We present the best sampled CO spectral line energy distribution (SLED) at $z = 6.003$, and analyzed it with the radiative transfer code MOLPOP-CEP. 
Fitting the CO SLED to a one-component model indicates a kinetic temperature $T_{\rm kin} = 228 \ \rm K$, molecular gas density $log (n(\rm H_{2})/\rm cm^{-3}$ )=4.75, and CO column density $log(N(\rm CO)/\rm cm^{-2}) =17.5$, although a two-component model better fits the data. In either case, the CO SLED is dominated by a "warm" and "dense" component. 
%A one-component model fit to the CO SLED indicates a ``warm" and ``dense" gas component with kinetic temperature $T_{\rm kin} = 228 \ \rm K$, molecular gas density $log (n(\rm H_{2})/\rm cm^{-3}$ )=4.75, and CO column density $log(N(\rm CO)/\rm cm^{-2}) =17.5$.
%Although a combination of a ``warm" component (that dominates the CO SLED at $J \geq 5$) and  a ``cold" component (that dominates the molecular gas mass but hardly contributes to the CO fluxes at $J \geq 5$) better fits the data.
Compared to samples of local (Ultra) Luminous Infrared Galaxies ((U)LIRGs), starburst galaxies and high redshift Submillimeter Galaxies (SMGs), J2310+1855 exhibits higher CO excitation at ($J \geq 8$), like other high redshift quasars.
The high CO excitation, together with the enhanced $L_{\rm H_{2}O}/ L_{IR} $, $L_{\rm H_{2}O}/ L_{CO} $ and $L_{OH^{+}}/L_{\rm H_{2}O} $ ratios, suggests that besides the UV radiation from young massive stars, other mechanisms such as shocks, cosmic rays and X-rays might also be responsible for the heating and ionization of the molecular gas. 
In the nuclear region probed by the molecular emissions lines, any of these mechanisms might be present due to the powerful quasar and the  starburst activity.

\end{abstract}


\keywords{galaxies: evolution – galaxies: high-redshift – galaxies: starburst – quasars: general – submillimeter: galaxies: general — radio line: galaxies}


% here to note for CO(X-X), use $CO (X-X)$ and also SDSS J231038.88+185519.7 should be $SDSS J231038.88+185519.7$
\section{Introduction} \label{introduction}
%Quasars are the most luminous sources in the early universe. These systems at the earliest evolutionary phase ($z \sim 6$) allow us to study the co-evolution of suppermassive black holes (SMBH) and their host galaxies since quasar-starburst systems have not yet had time enough to experience many mergers/secular evolution compared to the local samples (e.g. \citealt{walter04}; \citealt{targett12}; \citealt{wang10}).  
%Detailed studies on the molecular gas content in these high $z$ quasars greatly aid our understanding of the host and the SMBH co-growth, because the molecular gas content is the direct fuel of star formation in the host galaxy and also the material feeding the growth of the central SMBH. 
%Detections of bright [\ion{C}{2}]$\rm\ 158 \mu m$ and CO in these high $z$ systems reveal massive molecular gas reservoirs of $M_{\rm dyn}$ $\sim 10^{10} \ M_{\sun}$ within the central few kpcs (e.g., \citealt{maiolino05}; \citealt{wang10, wang11,wang16}; \citealt{riechers09}; \citealt{walter07}; \citealt{venemans17}; \citealt{carilli13}; \citealt{petric03}).

The quasars discovered at $z \gtrsim 6$ represent the first generation of super massive black holes (SMBHs) and host galaxies. 
Many among these earliest systems host SMBHs of $ \sim 10^{9}  M_{\sun}$ (e.g., \citealt{jiang07, jiang16}), and the strong dust continuum and [\ion{C}{2}] detections reveal dynamical masses of $ \sim 10^{10} - 10 ^{11}   M_{\sun}$ and star formation rate of $ \sim 10^{2} - 10^{3}  M_{\sun}\ \rm yr^{-1}$ in the host galaxies (e.g., \citealt{maiolino05};  \citealt{wang08, wang16,wang19}; \citealt{walter09}; \citealt{carilli13}; \citealt{venemans17b, venemans19}; \citealt{decarli18}; \citealt{neeleman19}).
These suggest that the SMBH and galaxy co-evolution is already in place in these $z \sim 6$ quasar-starburst systems.
In the meantime, bright molecular CO emission lines are widely detected in the starburst quasar hosts which reveal the molecular gas content of $ \sim 10 ^{9} - 10 ^{10}   M_{\sun}$ within a few kpc scale (e.g., \citealt{bertoldi03}; \citealt{walter03}; \citealt{riechers09}; \citealt{wang10,wang11, wang13, wang16}; \citealt{carilli13}; \citealt{venemans17a}).
In particular, the $z \gtrsim 6$ quasars are detected in very high (rotational quantum number) $J$  (e.g., $J \geq 9$) CO transitions, indicating high CO excitation comparable to that found in local extreme (Ultra)luminous Infrared Galaxys ((U)LIRGs) and AGNs (e.g., \citealt{gallerani14}; \citealt{wangf19}; \citealt{yang19}).
Taking the advantage of the most powerful sub-mm/mm and radio facilities, such as Atacama Large Millimeter/submillimeter Array (ALMA), NOrthern Extended Millimeter Array (NOEMA) and the Karl G. Jansky Very Large Array (VLA), extensive observations at submm/mm wavelengths have recently been carried out to search for the emission lines from the ionized, atomic, and molecular interstellar medium (ISM) (e.g., \citealt{decarli18}; \citealt{hashimoto18}; \citealt{walter18}; \citealt{novak19}). These observations are crucial for our understanding of the physical, chemical conditions and kinematics of the multi-phase ISM in these young quasar hosts at the earliest epoch and allow us to study the co-evolution of SMBHs and their host galaxies at the earliest evolutionary phase.

CO emission lines have long been the workhorse to probe the molecular gas at rest-frame sub-mm band in the local and high redshift Universe. 
Low $J$ (e.g., $J \lesssim 3$) CO emission lines are easy to excite in typical molecular cloud conditions (i.e., the lowest CO transition requires only $\sim$ 5 K above ground and densities of $\sim 100 \ \rm cm^{-3}$), thus it traces the bulk of the molecular gas content.
Mid $J$ (e.g., $4 \lesssim J \lesssim 8$) CO transitions are found to be linearly correlated with the far infrared luminosity and trace the star formation rate (e.g., \citealt{greve14}; \citealt{liu15}). UV photons from newly formed high-mass stars are responsible for the molecular gas heating at this regime.
The excitation of high $J$ (e.g., $J\gtrsim 9$) CO transitions require both high temperature and high density, which are usually related to the processes such as shocks, X-rays from AGN and cosmic rays ( \citealt{bradford03}; \citealt{spinoglio12}; \citealt{meijerink13};  \citealt{gallerani14}). % see Cloverleaf16 paper

CO spectral line energy distribution (SLED) $-$ the CO flux as a function of rotational quantum number $J$ is a probe of the molecular gas physical conditions (e.g., temperature, density and illuminating radiation field strength).
It has been used to study the physical conditions in a variety of local and high redshift systems (e.g., \citealt{weiss05, weiss07}; \citealt{riechers06}; \citealt{bradford09, braford11};  \citealt{spinoglio12};   \citealt{gallerani14}; \citealt{yang19}; \citealt{wangf19}). 
Normal star-forming galaxies have CO SLEDs that peak at relatively low $J$ CO transitions. E.g., the CO SLED of the Milky Way's inner disk peaks at around $J= 3 - 4$ \citep{fixsen99}, and similar CO SLEDs are found in other local star-forming galaxies \citep{daddi15}. 
Galaxies that experience higher star formation activity than normal galaxies, e.g., starburst galaxies, (U)LIRGs and submillimeter galaxies (SMGs), have moderately excited CO SLEDs which peak at higher $J$ compared to the star forming galaxies.
One of the closest examples is the local starburst galaxy M82, whose CO SLED peaks at $J = 5$ in the central region and show little line intensity at $J >9$ (\citealt{weiss05}; \citealt{panuzzo10}).
The molecular CO in luminous AGNs are usually highly excited. Very high $J$ CO transitions ($J\gtrsim 9$) are detected in well-known AGNs such as NGC 1068 \citep{spinoglio12}, Mrk231 \citep{van10}, the Cloverleaf quasar at $z = 2.56$ (\citealt{bradford09}; \citealt{uzgil16}) and the $z = 3.9$ lensed quasar APM 08279+5255 (\citealt{weiss07}; \citealt{riechers09}; \citealt{braford11}).
And known to date, bright CO emission lines at $J \geq 10$ were detected in the host galaxies of three $z\geq 6$ quasars, i.e.,  SDSS J114816.64+525150.3 (here after J1148+5251) at $z = 6.4$ (e.g., \citealt{riechers09}; \citealt{gallerani14}), SDSS J010013.02+280225.8 (here after J0100+2802) at $z = 6.3$ \citep{wangf19} and UHS J043947.08+163415.7 (hereafter J0439+1634) at $z=6.5$ \citep{yang19}.
The CO lines at $J \geq 10$ are likely to arise from warm gas with kinetic temperature of $T_{\rm kin} \geq 100$ K. Kinetic temperature is proportional to the kinetic energy through $T_{\rm kin} =\frac{2}{3} \frac{E_{n}}{k_{b}}$, where $k_{B}$ is Boltzmann constant and $E_{n}$ is the kinetic energy of the molecule.
In the high redshift quasars, X-rays from AGNs are frequently proposed to explain the CO excitation at high $J$ transitions (\citealt{bradford09}; \citealt{gallerani14};  \citealt{uzgil16}).

The  $\rm H_{2} O$ and $\rm OH^{+}$ lines provide additional diagnostics of heating and ionization source of the molecular gas (e.g., Cosmic rays, UV radiation, X-rays and shocks) in addition to CO.
The $\rm H_{2} O$ molecule traces the warm and dense molecular regions.
It is found to be bright in infrared luminous galaxies, and can even reach luminosities comparable to CO in these galaxies (\citealt{van11}; \citealt{omont13}; \citealt{yang13,yang16}; \citealt{ jarugula19}).
Recent studies found a nearly linear relation between the water  luminosity  and the infrared luminosity in local and high redshift systems over three orders of magnitude (\citealt{gonz10,gonz14};  \citealt{riechers13}; \citealt{yang13,yang17}; \citealt{omont13}; \citealt{jarugula19}). 
The brightest water lines are detected in the presence of shocks or X-rays (\citealt{gonz10}; \citealt{pellegrini13}).  
Accordingly, the $\rm H_{2} O$ emission  may even act as a tracer of the powering source of molecular gas, e.g., if the molecular gas is heated by the UV radiation or other mechanisms like shocks and X-rays. 
In addition, a variety of chemical processes are enrolled in the formation of $\rm H_{2} O$ and $\rm OH^{+}$.
The gas phase $\rm H_{2} O$ molecule is formed by either solid-phase or gas-phase chemical reactions (\citealt{pellegrini13}; \citealt{yang16}).
Neutral-neutral and ion-neutral reactions are two mechanisms to form $\rm H_{2} O$ molecule in the gas phase.
The former is usually related to the shocks, while the later is associated with PDRs, cosmic ray dominated regions, and X-ray dominated regions \citep{yang16}. 
The molecular ions (e.g., $\rm H_{2} O^{+}$, $\rm OH^{+}$) as intermediates of the ion-neutral reactions, play important roles in distinguishing between shocks and PDR/XDR/cosmic ray dominated region.
E.g., the bright $\rm H_{2} O^{+}$ line detection in high redshift lensed SMGs is probably initiated by cosmic rays \citep{yang16}.

The $\rm OH^{+}$ line, although not as strong as the turbulent gas tracer $\rm CH^{+}$ that has been detected in high redshift systems (e.g., \citealt{falgarone17}), traces the turbulent gas components as well, e.g., inflow or outflows. The formation of $\rm OH^{+}$ requires both atomic and molecular hydrogen, at which column density the cosmic-rays or X-rays are more capable of penetrating and ionizing the neutral and molecular gas (e.g.,\citealt{van10}; \citealt{meijerink11}; \citealt{gonz18}). The $\rm OH^{+}$ line has been detected both in absorption (that probes the cold turbulent gas) and in emission, where the chemical structure is dominated by the cosmic ray ionization or the X-ray radiation from the AGNs (e.g., \citealt{van10}; \citealt{gonz18}). Limited by the weak strength and the P Cygni line profile (presence of both absorption and emission in the profile of the same spectral line), there is only one reported $\rm OH^{+}$ line detection at $z \geq 6$  in absorption in the starburst galaxy HFLS3 \citep{riechers13}). 

In order to understand the physical conditions and the heating mechanisms of the ISM in the complex environment with both AGN and nuclear starburst activities in these young quasar hosts at $z \sim 6$, we here present a study of the CO SLED in one of the most far infrared and CO luminous quasars at $z \sim 6$, SDSS J231038.88+185519.7 (here after J2310+1855) at $z=6.003$.
J2310+1855 hosts a SMBH of $\approx$ 4 $\times \ 10^{9}\ M_{\sun}$ \citep{jiang16}. 
It was detected in bright dust continuum, CO $(2-1)$ and $(6-5)$, [\ion{C}{2}]$\rm \ 158\mu m$ and [\ion{O}{3}]$\ 88 \mu m$ lines (\citealt{wang13};  \citealt{do18}; \citealt{feruglio18}; \citealt{hashimoto18};  \citealt{carniani19}; \citealt{shao19}). 
This quasar is also detected in bright CO $(10-9)$ emission with a line flux of 1.04 $\pm$ 0.17 $\rm Jy \ km\ s^{-1}$ (Riechers et al. in prep).
The far infrared dust emissions suggest the host galaxy is actively forming stars with a star formation rate (SFR) of $\approx 2400 \  M_{\odot} \rm \ yr^{-1}$ , and is abundant in dust with a dust mass of $\approx 1.7 \times 10^{9}\  M_{\sun}$ (\citealt{wang13}; \citealt{hashimoto18}; \citealt{carniani19}; \citealt{shao19}). \citet{wang13} for the first time spatially resolved the large amount of gas ($\approx 9.6 \times 10^{10} \  M_{\sun}$) residing in a $0\farcs55$ $\times$ $0\farcs40$ ($\sim $ 3 kpc) disk based on the [\ion{C}{2}]$\rm \ 158\mu m$ observations. 
The CO $(2-1)$ emission line has a size comparable to the [\ion{C}{2}]$\rm \ 158\mu m$ emission with an associated molecular gas mass of $\approx 4.3 \times 10^{10} \  M_{\sun}$ \citep{shao17}.  
The [\ion{O}{3}]$\ 88 \mu m$ emission indicates a $L{\rm [O\ III]}$/ $L_{\rm IR}$ ratio comparable to local systems at similar $L_{\rm IR}$.
The quasar is luminous in X-ray as well and has a derived $L_{\rm 2-10kev} = 6.93 \times 10^{44} \rm erg\ s^{-1}$ \citep{vito19}. 
In addition, there is evidence of companions close to the quasar although further confirmation is needed (\citealt{do18}; \citealt{feruglio18}).
All together, J2310+1855 is an extraordinary quasar-starburst sample enabling us to study in detail a SMBH and galaxy co-evolution at $z \sim 6$.

In this paper, we present our new ALMA observations of CO $(8-7)$, $(9-8)$, $\rm H_{2}O (2_{0,2}-1_{1,1})$ and $ \rm OH^{+} (1_{1}-0_{1})$ and NOEMA observation of CO $(5-4)$, $(6-5)$,  $(12-11)$ and $(13-12)$ towards J2310+1855, aiming to investigate the molecular gas excitation mechanisms in environments of both intense star formation activity and luminous AGN.
The paper is organized as follows: in Section \ref{section2} and \ref{results}, we present the observations and results. In Section \ref{cosled mlodel}, we analyze the CO SLED with a radiative transfer code to probe the physical conditions of molecular gas. In Section \ref{discussions}, we compare the CO excitation in J2310+1855 with local and high redshift galaxy samples and AGNs and discuss the heating mechanisms of molecular gas as well. Finally, we summarize the results in Section \ref{conclusions}. We adopt a flat $\Lambda$CDM cosmology with $H_{0}=70\ \rm km \ s^{-1}\ Mpc^{-1}$ and $\Omega_{\rm m}=0.3$, where 1$''$ corresponds to 5.7 kpc at the J2310 + 1855 redshift ($ z=6.0031$), and the luminosity distance to J2310+1855 is 57763 Mpc.
\section{Observations} \label{section2}  
\subsection{ALMA}
We observed the CO $(8-7)$ ($\nu _{\rm rest}$= 921.7997 GHz), CO $(9-8)$ ($\nu _{\rm rest}$= 1036.9124 GHz), $\rm H_{2}O (2_{0,2}-1_{1,1})$ ($\nu _{\rm rest}$= 987.9268 GHz, hereafter $\rm H_{2}O$) and $\rm OH^{+} (1_{1}-0_{1})$ ($\nu _{\rm rest}$= 1033 GHz, hereafter $\rm OH^{+}$)
 emission lines,  as well as the underlying continuum towards $z = 6.003$ quasar J2310+1855 with ALMA (Cycle 3, ID 2015.1.01265.S). All the observations were executed between April and November in 2016 with beam sizes between $0\farcs6$ and $0\farcs8$. 
The CO $(8-7)$, $(9-8)$, $\rm H_{2}O$ and $\rm OH^{+}$ lines were observed in ALMA Band 4 with two separate executions, where 36 to 44 12-m diameter antennas were used during observations. For each observation, we used four spectral windows, each with a width of 1.875 GHz consisting of 128 channels, with two of the windows in the lower sideband (LSB) and the other two in the upper sideband (USB). 
The CO $(8-7)$ and $\rm H_{2}O$ lines were observed in one spectral setup, with one spectral window centered on the CO $(8-7)$ observed frequency of 131.6274 GHz, one window covering the $\rm H_{2}O$ emission, and the other two covering line-free dust continuum.
In the other turning, we observed the CO $(9-8)$ line centered at the frequency of 148.0648 GHz with the $\rm OH^{+}$ line also covered in the same spectral window, while the other three windows measured the dust continuum. 
The fluxes were calibrated using the standard flux calibrator $Pallas$, while SDSS J2253+1608 was used as both the phase calibrator and the bandpass calibrator. The typical calibration uncertainty is $< 5\%$ in ALMA band 4, here we use $ 15\%$ uncertainty that also includes the uncertainties of  the old $Pallas$ flux model in early casa versions \citep{stanley19}.

The observational data were calibrated and reduced with the Common Astronomy Software Applications (CASA) software package version 4.7.0 \citep{casa07}, using the standard ALMA pipeline. 
The maps were generated using the CLEAN task in CASA, and we apply the robust weighting algorithm with a Briggs parameter of 2 equivalent to natural weighting.  This results in a FWHM synthesized beam size of $0\farcs79$ $\times$ $0\farcs75$ and $0\farcs77$ $\times$ $0\farcs63$ at CO $(8-7)$($\rm H_{2}O$) and CO $(9-8)$($\rm OH^{+}$) observing frequency, and $0\farcs75$ $\times$ $0\farcs72$ and $0\farcs80$ $\times$ $0\farcs65$ at 136.6 and 141.1 GHz for the continuum.
The continuum level was determined using a first order polynomial, and the emission lines were imaged from the continuum subtracted data cube with all the line emitting channels included. With a total on source time of 34.9 and 30.3 minutes for CO $(8-7)$($\rm H_{2}O$) and CO $(9-8)$($\rm OH^{+}$), we finally binned the data of CO $(8-7)$($\rm H_{2}O$) and CO $(9-8)$($\rm OH^{+}$) to 36 and 32 $\rm km\ s^{-1}$, and the corresponding rms sensitivities were 0.17 and 0.19 $\rm mJy\ beam^{-1}$ respectively. The sensitivity of the underlying continuum was 15 $\rm \mu Jy\ beam^{-1}$.

\subsection{NOEMA}
We observed CO $(5-4)$ ($\nu _{\rm rest}$=  576.2679 GHz), CO $(6-5)$ ($\nu _{\rm rest}$= 691.4731 GHz), CO $(12-11)$ ($\nu _{\rm rest}$= 1381.9951 GHz), CO $(13-12)$ ($\nu _{\rm rest}$= 1496.9229 GHz) and the underlying continuum of this quasar with NOEMA (Project W18EE).
The CO $(5-4)$ and $(6-5)$ lines were observed in one tuning with the PolyFix correlator in Band 1 (3 mm), with the CO $(5-4)$ line in the LSB and CO $(6-5)$ in the USB, each with 7.744 GHz bandwidth.
The observations were executed in A configuration on 2019 Jan 18 with a total observing time of 2 hours, with 1.18 hours on source while the rest of the time was expended for calibrations. A total number of 8 or 9 antennas were used. 
The CO $(12-11)$ and $(13-12)$ lines were observed in C/D configuration in Band 3 (1 mm) with one frequency setup, with the CO $(12-11)$ line in the LSB and  CO $(13-12)$ in the USB, each with a 7.744 GHz bandwidth. 
The observations started on 2019 Apr 17 and ended on 2019 May 1. The total observing time was 8 hours with 6.2 hours on source, and a total number of 8 to 10 antennas were used in the observations.
3C454.3 was used as phase calibrator throughout all the CO observation. The typical calibration uncertainty is $< 10 \%$ in the 3 mm band and $< 20 \%$ in the 1 mm band. 
%see neri_noema.pdf

The data were reduced with the Grenoble Image and Line Data Analysis System (GILDAS) software \citep{gui00} packages CLIC and MAPPING. 
We extracted the continuum from all line free channels in the $uv$ plane with UV$\_$AVERAGE.
The $uv$ table of spectral lines was generated through UV$ \_$SUBTRACT with the underlying continuum subtracted.
Both the $uv$ table of the continuum and spectral lines were cleaned with the HOGBOM algorithm and NATURAL weighting was used to ensure the maximum S/N.
This results in a FWHM synthesized beam size of  $1\farcs67$ $\times$ $1\farcs37$/$1\farcs42$ $\times$ $1\farcs19$  for CO $(5-4)$/$(6-5)$ and $2\farcs08$ $\times$ $1\farcs62$/$1\farcs91$ $\times$ $1\farcs53$ for CO $(12-11)$/$(13-12)$.
The calibrated data of CO $(5-4)$ and $(6-5)$ were smoothed by a factor of 8 in frequency, resulting a spectral resolution of 16 MHz ($\sim$ 60 $\rm km\ s^{-1}$), and the calibrated CO $(12-11)$ and $(13-12)$ data were binned to 40 MHz ($\sim 60\ \rm km\ s^{-1}$) resolution.
The sensitivity was 0.32 $\rm mJy\ beam^{-1}$ for CO $(5-4)$ and $(6-5)$ and 0.54 $\rm mJy\ beam^{-1}$ for CO $(12-11)$ and $(13-12)$ per binned channels.

The observational details are listed in Table \ref{table1}.

\section{Results} \label{results}
With ALMA, we detect the CO $(8-7)$, $(9-8)$ and $\rm H_{2}O$ emission lines at 25$\sigma$,  22$\sigma$, and  15$\sigma$, respectively. The CO emission lines and the $\rm H_{2}O$ line are marginally resolved. We also obtained a tentative signal (4$\sigma$) for the $\rm OH^{+}$ line on the red wing of the CO $(9-8)$ line.  The line intensity maps integrated over the line emitting channels are presented in Figure \ref{figure1}. The right panel of Figure \ref{figure1} shows the line spectra integrated within the 2$\sigma$ contour in the intensity map.   
We calculate the line widths, redshift, and fluxes of CO  $(8-7)$, $(9-8)$ and $\rm H_{2}O$ by fitting a Gaussian profile to the spectra. We fit a 2D Gaussian component to the intensity maps of CO and $\rm H_{2}O$ lines, and the source sizes are derived by deconvolving the fitted component with the beam.
The spectral profiles of CO $(9-8)$, $(8-7)$ and $\rm H_{2}O$ are similar (Figure \ref{figure7}), suggesting that the high $J$ CO and $\rm H_{2}O$ lines are probing similar regions.
The source sizes measured from CO $(9-8)$, $(8-7)$ and $\rm H_{2}O$ lines of $\approx$ (0\farcs 4 $\pm$ 0\farcs 1) $\times$ (0\farcs 3 $\pm$ 0\farcs 1) are sightly smaller than values found from previous CO $(2-1)$ observation of (0\farcs 6 $\pm$ 0\farcs 2) $\times$ (0\farcs 4 $\pm$ 0\farcs 2) and  [\ion{C}{2}]$\rm \ 158\mu m$ observation of (0\farcs 6 $\pm$ 0\farcs 1) $\times$ (0\farcs 4 $\pm$ 0\farcs 1) at similar spatial resolution (\citealt{wang13}; \citealt{feruglio18}; \citealt{shao19}).
This may imply that CO $(9-8)$, $(8-7)$ and $\rm H_{2}O$ lines trace similar dense molecular regions that are closer to the central SMBH compared to CO $(2-1)$ and  [\ion{C}{2}]$\rm \ 158\mu m$.
The redshift measured with the CO $(9-8)$, $(8-7)$, and $\rm H_{2}O$ lines are within the uncertainties consistent with that from previous [\ion{C}{2}],  CO $(2-1)$  and $(6-5)$ observations (\citealt{wang13}; \citealt{feruglio18}; \citealt{shao19}). 
The  $\rm OH^{+}$ line is not as strong as the other detections, so we fix the center frequency to the [\ion{C}{2}]$\rm \ 158\mu m$ redshift, and fit a Gaussian profile to the spectra extracted from the peak pixel. 
As for the line widths, all the ALMA detections show line widths of $\sim$ 400 $\rm km \ s^{-1}$ consistent with previous CO and [\ion{C}{2}] observations. 
From the $\rm OH^{+}$ spectra, we find that there is an absorption like feature in the line center frequency, but the current S/N ratio is insufficient to confirm this feature.
The continuum detection were published in \citet{shao19}. The derived continuum source sizes of $\approx$ (0\farcs 30 $\pm$ 0\farcs 04) $\times$ (0\farcs 22 $\pm$ 0\farcs 06) are comparable to that measured with the CO and $\rm H_{2}O$ lines.

We detected CO $(5-4)$, $(6-5)$, $(12-11)$ and $(13-12)$ with NOEMA. All of the four CO lines are unresolved. For the  line widths, redshift, and fluxes calculation, we fit a Gaussian profile to the spectra extracted from the peak intensity pixel.
The redshift measured with the CO $(5-4)$, $(6-5)$, $(12-11)$ and $(13-12)$ are consistent with our ALMA detections as well as previous CO and [\ion{C}{2}] detections.  
The line width detected in ALMA and NOEMA observations are consistent.
The CO $(6-5)$ line has been previously detected with ALMA  and the (pre-NOEMA) IRAM Plateau de Bure Interferometer (PdBI) with different spatial resolutions and the spectral line flux measured with ALMA is found to be only 70$\%$ of that found with the PdBI (\citealt{wang13}; \citealt{feruglio18}). 
In addition to the calibration uncertainties in different observations, it is also possible that the low resolution PdBI data include more flux from the extended region.
The new NOEMA observation yield a CO $(6-5)$ flux of 1.05 $\pm$ 0.07 Jy $\rm km \ s^{-1}$, consistent with the results obtained with ALMA \citep{feruglio18}. 
We also detected the underlying continuum at high S/N ratio.
The measurements of redshift, line widths, line fluxes, and deconvolved source sizes of our ALMA and NOEMA observations as well as previous detections are summarized in Table \ref{table2}, and the continuum measurements are listed in Table \ref{table3}. 
The continuum, line intensity maps and spectra of CO $(5-4)$, $(6-5)$,  $(8-7)$, $(9-8)$, $(12-11)$, $(13-12)$,  $\rm H_{2}O$ and $\rm OH^{+}$  are shown in Figure \ref{figure2} (NOEMA) and  \ref{figure1} (ALMA). 
%We  find that all the ALMA detected intensity and continuum peaks are slightly spatially offset from the Gaia position of the quasar {\color{red} add values and uncertainties}. While the peak offset observed in NOEMA is within uncertainties.

% Thermal noise is 1/2*beam/SNR  SNR 22 19 15 for ALMA 0.65, 0.62, 0.6  resulting thermal 0.015  0.016  0.02
% Thermal noise is 1/2*beam/SNR  SNR 10 14 11 5 for NOEMA 1.67 1.42, 2.08, 1.91  resulting thermal 0.08  0.051  0.095 0.19

Figure \ref{figure3} shows the velocity and velocity dispersion maps of [\ion{C}{2}]$\rm \ 158\mu m$ \citep{wang13}, CO $(8-7)$, $(9-8)$ and $\rm H_{2}O$. 
The velocity fields of CO $(8-7)$, $(9-8)$ and $\rm H_{2}O$ overall follow the velocity gradient observed in [\ion{C}{2}]$\rm \ 158\mu m$ from north to south, which indicates that the emission might trace a rotating molecular gas disk. 
%(e.g., with $0\farcs076$ resolution [\ion{C}{2}] observation, \citet{venemans19} reported some non-rotating structures that were not observed by previous observations at lower resolution). 
The velocity field of CO $(8-7)$ shows a high velocity dispersion part in the western part, which is not observed in CO $(9-8)$ and $\rm H_{2}O$.
Such irregular velocity structure in CO $(8-7)$ is likely to be a result of the low S/N.
As for the velocity dispersion, CO $(8-7)$, $(9-8)$ and $\rm H_{2}O$ show velocity dispersion of $\textless 100 \rm \ km \ s^{-1}$ in the outskirts (that is not likely influenced much by the beam smearing effect).
 Higher S/N observations, possibly at even higher angular resolution, are required in order to constrain the kinematic structures of the dense molecular gas. 

\section{Radiative Transfer Analysis of the CO SLED} \label{cosled mlodel}
In the CO SLED analysis, we also include a new detection of the CO (10-9) line from NOEMA  at high S/N ratio. More details of the observation will be described in Riechers et al. (in prep). 
Our ALMA and NOEMA data, together with previous detections of the CO $(2-1)$ \citep{shao19} and the CO $(10-9)$ emission line, enable us to probe the the CO SLED of J2310+1855 from $J=2$ to $J=13$, making it the most complete CO SLED ever obtained for a $z \gtrsim 6$ quasar.
We here use the radiative transfer model MOLPOP-CEP to investigate the physical conditions of the molecular gas,  including the kinetic temperature $T_{\rm kin} $, molecular hydrogen density $n(\rm H_{2}) $ and CO column density $N(\rm CO)$.

%We will use two models to study the physical conditions of the molecular gas through the CO SLED analysis, including 1) a model consisting of one gas component with physical parameters of ($T_{\rm kin} $, $n(\rm H_{2}) $, $N(\rm CO) $), 2) a model consisting of two gas components but with different kinetic temperature $T_{\rm kin} $, density $n(\rm H_{2}) $ and CO column density $N(\rm CO) $.
%This approach has several advantages:
%(a) The escape probability is derived rigorously from the first principle, which makes it an exact method, 
%(b) Dividing the geometry into several zones makes it possible to solve the level populations as a function of depth into the emission region and also leads to more accurate solutions compared to previous large velocity gradient (LVG) models,
%(c) The computational time is rather short given its complexity.
%These two different treatments compared to previous codes that solves level population equations e.g. Radex, makes very different predictions for the escape velocity at fixed optical depth. Thus one has to be careful when comparing the results of this models with previous codes.
%These advantages make it an exact and fast method to predict the line intensities with given physical parameters.

\subsection{Method}    \label{method}
MOLPOP-CEP is a universal code that enables exact solutions of multi-level line emissions radiative transfer problems
for all the atoms/molecules that have atomic/molecular data in the Leiden Atomic and Molecular Database (LAMDA) database  (\citealt{elitzur06}; \citealt{asensio18}). 
This code assumes a slab geometry with the emitting region divided into several zones and treats the radiative transfer problem with a coupled escape probability (CEP) method that aims to solve the coupled level population equations of different zones under consideration.
Dividing the geometry into several zones makes it possible to solve the level populations as a function of depth into the line emitting region and also leads to more accurate solutions compared to previous large velocity gradient (LVG) models.

The physical parameters as inputs control the number of zones that the geometry is divided and the physical parameters of the individual zone in the slab geometry. 
Even for a uniform physical parameter setup in the whole geometry, the division of geometry into zones is necessary to increase the accuracy of the results. This is because, for optically thick lines, the strength of radiative reactions changes with distances to the surface, and the transition level population distributions depend on positions in the geometry \citep{asensio18}. 
In the slab, each zone in principle can have different physical parameter setups including: (1) the zone width $\Delta L$, (2) the gas density within the zone $n(\rm H_{2}) $, (3) kinetic temperature $T_{\rm kin} $, (4) molecular abundance, (5) local linewidth (which corresponds to the line absorption/emission profile in each point in the geometry). 
Besides these, MOLPOP-CEP allows the inclusion of external radiation field. 
Given these physical parameters, the code will then solve the coupled level population problem between zones and finally predict the emergent intensities of the emission lines that can be directly compared to the observations. 
We assume uniform parameters for each slab.
The accuracy of MOLPOP-CEP solutions increases with the number of zones. Here we divide the geometry into 10 zones for the model calculation as is suggested by \citet{asensio18}. 

We generate a grid of slab models through varying the physical parameters of greatest interest $-$ the gas density $n(\rm H_{2}) $, temperature $T_{\rm kin} $ and zone width $\Delta L$. For the other two zone parameters, we fix the molecular abundance of CO to $X_{\rm CO} = 10^{-4}$ (Milky Way, \citealt{blake87}),  and local linewidth to 1 $\rm km\ s^{-1}$ for all the model calculations.
In addition, we include the CMB at the quasar redshift of 19.12 K, because the hot CMB at high redshift will 1) acts as an extra heating source of the CO emission, 2) serves as a continuum background (\citealt{dacunha13}; \citealt{zhang16}).
In each model, all of the 10 zones have the same physical parameter setup thus each model in the grid can be described by a uniform $n(\rm H_{2}) $, $T_{\rm kin} $ and $\Delta L$.
The CO column density within the geometry, which is the sum of the column densities of 10 zones, is proportional to the zone width $\Delta L$ through:
\begin{align}
 N(\rm CO) = 10 \times \textit{n}(\rm H_{2}) \times \Delta \textit{L} \times \textit{X}_{\rm CO}.
%\label{eq200}
\end{align} 
In the rest of the paper, we use $n(\rm H_{2}) $, $T_{\rm kin} $ and $N(\rm CO)$  to characterize the physical condition of each slab grid. As it is easier to use $N(\rm CO)$ rather than $\Delta L$ to make comparisons with constraints from observations (see section \ref{parameter constraints}).
The final grid covers the typical physical conditions of the molecular clouds with a temperature range of $\rm 20 -800\ K$, density of $\rm 10^{3} - \rm 10^{8} \ \rm cm^{-3}$ and CO column density of $\rm 10^{14} - \rm 10^{21} \ \rm cm^{-2}$ (this corresponds to  $\rm H_{2}$ column density of $\rm 10^{18} - \rm 10^{25} \ \rm cm^{-2}$ for our assumed $X_{\rm CO}$ of $10^{-4}$). More details about the grid are listed in Table \ref{table4}.

We use the model grid to fit our observed CO SLED of J2310+1855. The fitting procedure is as follows: we first apply the least square method to find the best fitting results. We also use the Bayesian code emcee \citep{foreman13} to efficiently explore the parameter space and get the posterior probability distributions of all the parameters considered.
Emcee is an extensible, pure-Python implementation that is designed for Bayesian parameter estimation using Ensemble samplers with affine invariance \citep{jonathan10}. 
%With this Bayesian analysis, we are modeling three physical parameters ($T_{\rm kin}$, $n(\rm H_{2})$, $N(\rm CO)$) simultaneously to the observed CO SLED, and are able to get the posterior probability distributions of all the parameters considered.

\subsection{Parameter Constraints}\label{parameter constraints}
During the fitting procedure, we set constraints for the three parameters: $T_{\rm kin} $, $n(\rm H_{2}) $ and $N(\rm CO) $. 
The fact that $T_{\rm kin} $ is hotter than the background CMB  radiation at redshift 6 is a prior, which sets $T_{\rm kin} >$19.12 K. 
As for column density, one prior is that the total amount of gas producing the observed CO luminosities should be no more than the total dynamical mass of this system, this leads to:
\begin{align}
N(\rm CO) \Phi_{\textit {A}} < \frac{\textit{M}_{\rm dyn} \textit{X}_{\rm CO}}{\mu \textit{m}_{\rm H2} \textit{A}}\ [\rm cm^{-2}],
\label{eq7}
\end{align}
% references schirm17
where $\Phi_{\textit {A}}$ is the filling factor, $M_{\rm dyn}$ is the dynamic mass, $A$ is the source area in $\rm cm^{2}$, $\Delta V$ is the CO line width in $\rm km\ s^{-1}$, $\mu$ is the mean molecular weight, $m_{\rm H2} $ is the $\rm H_{2}$ molecule mass and $X_{\rm CO}$ is the CO abundance. The source size and gas dynamical mass are adopted from \citet{shao19}, with $M_{\rm dyn} \approx 4.3 \times 10^{10}\ M_{\sun}$ and $A \approx \rm \frac{\pi}{4} \times 0.60 \times 0.40\ arcsec^{2}$.
Adopting a CO to $\rm H_{2}$ abundance ratio $X_{\rm CO} = 10^{-4}$, and $\mu = 1.4$ for mean molecular weight, the final constraint is:
\begin{align}
N(\rm CO) \Phi_{\textit {A}} < 3.2 \times 10^{19} \ \rm [cm^{-2}].
\label{eq8}
\end{align}
The other prior for column density is that it should be less than the source gas volume density integrated along line of sight for line emitting regions. This leads to: 
\begin{align}
N(\rm CO)  < \textit{n}(\rm H_{2}) \times \textit{X}_{ CO} \times \textit{S}\ [cm^{-2}],
\label{eq10}
\end{align}
where $S$ is the source size along  the line of sight, $n(\rm H_{2})$ is the $\rm H_{2}$ volume density. 
Assuming a largest diameter of CO emitting region of $0\farcs60$ \citep{shao19}, we set a constraint on both $ N(\rm CO)$ and $n(\rm H_{2})$ as follows:
\begin{align}
\frac{N(\rm CO) }{n(\rm H_{2})} <  1.1 \times 10^{18} \rm [cm].
\label{eq11}
\end{align}

\subsection{Fitting Results} \label{fitting results}
We endeavor to probe the physical conditions of the molecular gas in J2310+1855 by fitting a one-component model to the CO SLED. Calibration uncertainties are included in all the modeling processes throughout the paper.
We first fit the grid models to the observed CO SLED with the least square method. 
The left panel of Figure \ref{figure4} shows the best fitting result, suggesting a ``warm" and ``dense" gas component with kinetic temperature of $T_{\rm kin} = 228 \ \rm K$, density of $log (n(\rm H_{2})/\rm cm^{-3} )=4.75$ and column density of $log(N(\rm CO)/\rm \ cm^{-2}) =17.5$. 
We then search for all the possible physical conditions that fit the observational CO SLED with the emcee code. The posterior probability distributions of the three parameters are shown in Figure \ref{figure5}. 
This indicate that the data can be fitted with a ``warm" and ``dense" gas component with parameter range of $T_{\rm kin} \approx 167^{+153}_{-56} \ \rm K$, $log (n(\rm H_{2}) / \rm cm^{-3})\approx 5.11^{+1.83}_{-0.58}$ and $log(N(\rm CO)/\rm \ cm^{-2}) \approx 17.28^{+0.33}_{-0.42} $. {\footnote{The resulting parameter ranges are consistent within 1$\sigma$ between including and excluding the CO $(10-9)$ line in the fitting procedure.} }
But the current best fit model fail to reproduce the very high CO $(8-7)$ line flux detected.

Previous CO SLED modeling from local to high redshift galaxies/AGNs suggest different gas physical properties in different systems. E.g., in the $z=2.56$ quasar cloverleaf \citep{riechers11} and the $z=6.34$ starburst galaxy HLFS3 \citep{riechers13}, a single gas component is able to reproduce the observed CO SLED.
In addition, more than one gas component is found in the CO SLED analysis of local starburst galaxies, (U)LIRGs, and even the quasars at the highest redshift (e.g.,  M82, NGC 1068, Mrk 231, APM 08279+5255, J0100+2802, J0439+1634 and J1148+5251 (\citealt{weiss05, weiss07};  \citealt{panuzzo10}; \citealt{van10}; \citealt{spinoglio12}; \citealt{gallerani14}; \citealt{wangf19};  \citealt{yang19})).
The CO SLED of the $z = 6.3$ quasar J0100+2802 suggests two components of gas with a ``cold" component with $T_{\rm kin} \approx$ 24 K, $log (n(\rm H_{2})/ \rm cm^{-3} ) \approx4.5$ and a ``warm" component with  $T_{\rm kin} \approx$ 224 K, $log (n(\rm H_{2}) /\rm cm^{-3}) \approx3.6$ \citep{wangf19}.
The CO SLED of J0439+1634 at $z=6.5$ indicates a ``cold" component with $T_{\rm kin} \approx$ 23 K and $log (n(\rm H_{2})/ \rm cm^{-3} ) \approx 4.1$ in combination with a ``warm" component with $T_{\rm kin} \approx$ 140 K and $log (n(\rm H_{2}) / \rm cm^{-3}) \approx 4.5$ \citep{yang19}.
The cold component was thought to be associated with the submm/mm-detected dust component powered by active star formation with temperatures of 40 $\sim$ 60 K (assuming optically thin, \citealt{beelen06}; \citealt{leipski13}). 
The dust continuum SED fitting  of J2310+1855 indicates a dust temperature of  $\sim$ 40 K in the optically thin dust assumption of \citet{shao19} or 76 K in the optically thick assumption of \citet{carniani19}. Both are much lower than the one-component fitting result of  $T_{\rm kin}$ = 228 K. 
As the one-component best-fit model fails to explain the observed CO (8-7) flux (left panel of Figure \ref{figure4}), it is possible that there is an additional ``cold" gas component physically associated with the submm/mm detected dust in J2310+1855.
Motivated by the above, we examine whether the data can be explained with a two-component model. 

The two-component model fitting to the data requires eight parameters, including the physical parameters ($T_{\rm kin} $, $n(\rm H_{2}) $ and $N(\rm CO) $) and the normalization of each component. Our data are insufficient to constrain all eight parameters. 
Considering that the "cold" component is usually physically connected to and has similar temperature as that of the cold dust (as is explained in detail in the previous paragraph), and the dust temperature of J2310+1855 is not well constrained ($T_{dust}$ ranges between 40 $-$ 80 K depending on the dust model assumed), we will fix the "cold" component to the typical "cold" gas physical conditions observed in z $\sim$ 6 quasar in the following analysis. With this assumption, we are fitting only five instead of eight parameters. As a consequence, the model parameters could be better constrained. We adopt a typical "cold" gas component with physical parameters of $T_{\rm kin} \approx$ 50 K and $log (n(\rm H_{2})/ \rm cm^{-3} )=4.2$ that is observed in a "typical" $z\sim 6$ quasar J1148+5251 \citep{riechers09}.
Because column density is not one of the model parameters in \citet{riechers09}, we use a column density of $log(N(\rm CO)/\rm \ cm^{-2}) =18.0$ for the ``cold" model. We note that the resulting $X_{CO} / dv /dr$ here is different from that in \citet{riechers09}, because in the MOLPOP-CEP model, the $dv/dr$ is derived from the first principle and is different in different places in the whole geometry. 
The final ``cold" model using the set of parameters we adopted here can well represent the observational CO SLED of J1148+5251 presented in \citet{riechers09}.
The right panel of Figure \ref{figure4} shows the minimum $\chi^{2}$ fitting result of the two-components model. The resulting ``warm" component with the minimum $\chi^{2}$ has a physical condition of $T_{\rm kin} = 306 \ \rm K$, $log (n(\rm H_{2})/\rm cm^{-3} )=5.25$ and $log(N(\rm CO)/\rm cm^{-2}) =15.5 $. We find in the fitting result that the ``cold" component (the J1148+5251 model) dominates the low $J$ ($J=2$) part, and contributes to $77 \%$  of the observed CO $(2-1)$ flux. 
As we have already mentioned before, the low $J$ CO emission lines trace the total molecular gas mass, thus the ``cold" component dominates the total molecular gas mass. %contributes at least to $77 \%$ of the total molecular gas mass (this is because taking into account the CMB effect, this fraction becomes $84 \%$). 
In the mid-$J$ ($J=5,\ 6$) part, the contribution of the ``cold" component decreases and only accounts for $\sim 30\%$ of the observed CO fluxes. And in the high $J$ ($J\geq 8$) part, the ``cold" component contribution is negligible.  The ``warm" component, that barely contributes to the total molecular gas mass, dominates the CO SLED from the mid- to the high-$J$ ($J \geq 5$) part of the overall CO SLED.
The posterior probability distributions of the parameters calculated by the emcee code are shown in Figure \ref{figure5}. It suggest a ``warm" and ``dense" component with parameter range of $T_{\rm kin} \approx 306^{+263}_{-149}\  \rm K$, $log (n(\rm H_{2}) / \rm cm^{-3}) \approx 5.22^{+1.04}_{-0.49}$ and $log(N(\rm CO)/\rm \ cm^{-2}) \approx 15.29^{+1.34}_{-1.17}$. {\footnote{The resulting parameter ranges are consistent within 1$\sigma$ between including and excluding the CO $(10-9)$ line in the fitting procedure.} }

To summarize, the best one-component ``warm" and ``dense" model reproduces the observed CO SLED in general, except for an underestimation of the CO $(8-7)$ flux.
The two-component fitting result suggests the CO SLED at $J \geq 5$ is dominated by a ``warm" and ``dense" gas component, while the ``cold" component barely contribute to the mid- to the high-$J$ CO fluxes but dominate the total molecular gas mass. 
Either one or two component model suggests that the CO SLED detected within the nuclear region (source size of $\sim \ 2 \ \rm kpc$) of the quasar host is dominated by a ``warm" and ``dense" gas component at $J \geq 5$. 
%And we are not able to rule out a ``cold" component with temperature comparable to the cold dust that dominates the total molecular gas mass but hardly contributes to CO fluxes at $J \geq 5$.

\section{Discussion} \label{discussions}
\subsection{The CO emission in J2310+1855 compared with local starburst systems}\label{localsb}
The CO SLED reveals the physical conditions of molecular gas (e.g., the illuminating radiation field strength, kinetic temperature, volume density and column density). We first compare the CO SLED of J2310+1855 with local starburst systems.
Figure \ref{figure6} (a) shows the CO SLED of J2310+1855  compared with two local starburst samples. They are local (U)LIRGs sample consisting of 29 (U)LIRGs \citep{rosen15} and local normal + starburst galaxy sample consisting of 43 star-forming galaxies (SFGs) and 124 (U)LIRGs \citep{liu15}. 
Although there are actually a small number of AGNs contained in the two (U)LIRGs comparison samples, they are confirmed to impact negligibly on both the CO flux and the infrared luminosity  \citep{rosen15}. For the  \citet{rosen15} sample, we exclude NGC 6240 in sample mean calculation (because this AGN represents a very extreme CO SLED, see details in Section \ref{localagn}).
Through comparisons, we find that the peak of CO SLED is $\lesssim 4$ for all the local starburst samples, while the J2310+1855 CO SLED peaks at much higher $J$ transitions at $J=8$. The CO emission lines of J2310+1855 show higher excitation compared to all local (U)LIRG samples (i.e., the CO flux is higher than the average of all the local ULIRG samples and is also well above the range of all the comparison samples especially for $J\geq 8$).
In addition, we compare J2310+1855 with a representative example of local starburst galaxy M82 in Figure \ref{figure6} (b). We get a similar result  as that of the starburst samples. The CO SLED of M82 peaks at $J=5$ and decreases dramatically at $J \geq 8$,  contrary to J2310+1855 that peaks at $J=8$ and is luminous even at $J \geq  10$.

The high CO excitation detected in the J2310+1855 nuclear region (source size of $\sim$ 2 kpc in FWHM) may indicate other heating mechanisms besides the UV heating from massive young stars (e.g., mechanical heating by shocks, X-ray heating from AGNs), or very intense UV radiation field (e.g., large UV photon flux produced by a result of both the quasar and the star formation, see Section \ref{mechanisms} for further discussions about these heating mechanisms).

\subsection{The CO emission in J2310+1855 compared with local AGNs}\label{localagn}
We compare the CO emission lines of J2310+1855 with some representative local AGNs (Mrk 231, NGC 1068 and NGC 6240). The AGNs we selected are thoroughly studied local AGNs that represent different CO heating mechanisms. %, including the XDR heating of the high $J$ CO in the very nuclear region (Mrk 231 and NGC 1068), and the mechanical heating of high $J$ CO by shocks (NGC 6240).
NGC 1068 is one of the closest AGN, whose high $J$ CO emission lines in the circumnuclear disk (CND) are best explained by a XDR model, and the starburst ring at larger radii that dominates the molecular gas mass is best fitted with a PDR model \citep{spinoglio12}. Mrk 231 requires an XDR model to fit the high $J$ CO emission lines in the central 160 pc molecular region, although the low $J$ CO lines are mainly from the PDR component at larger distances \citep{van10}. NGC 6240 is a local AGN with three nucleus, the CO emission lines are unlikely to correlate with the position of either AGN nuclei and mechanical heating is proposed to interpret the extremely excited CO SLED and also the optical ISM emission lines (\citealt{meijerink13}; \citealt{kollatschny19}). 

Figure \ref{figure6} (b) shows the CO SLED of J2310+1855 and the local AGNs. The CO SLED of NGC 1068 peaks at $J \leq 4$, and then decreases rapidly with increasing $J$. Mrk 231 shows an extreme CO SLED that peaks at $J=5$ and displays a high normalized CO flux even at $J \approx 10-13$. The differences between  NGC 1068 and Mrk 231 can be explained by different X-ray energy input to the CO heating,  evident from the higher X-ray flux derived in Mrk 231 (28 erg $\rm cm^{-2}\ s^{-1}$, \citealt{van10}) than NGC 1068 (9 erg $\rm cm^{-2}\ s^{-1}$, \citealt{spinoglio12}). NGC 6240 shows the most extreme CO SLED among these three AGNs. 
The quasar J2310+1855 peaks at higher $J$ ($J=8$) compared to these local AGNs, i.e., the gas in the nuclear region of J2310+1855 has higher excitation compared to that of the local AGNs.

\subsection{The CO emission in J2310+1855 compared with high redshift systems}
We also compare the CO SLED of J2310+1855 with high redshift systems, including high redshift (lensed) SMGs and quasars.
Figure \ref{figure6} (a) shows the CO SLED of J2310+1855 and high redshift SMGs: a sample of $z \sim 1.2-4.1$ SMGs \citep{bothwell13} and a sample of 15 $z \sim 2-4$ lensed SMGs from \citet{carilli13} and \citet{yang17}.
The high redshift SMG CO SLEDs peak at $J \lesssim 6$, while J2310+1855 peaks at higher $J$ ($J=8$) than the two SMG samples.
This is similar to the results when comparing the J2310+1855 CO SLED to local starburst samples.

We also select some well-known high redshift quasars for comparison. Including two lensed quasars, the Cloverleaf at $z=2.56$ and APM 08279+5255 at $z=3.91$. The fit to the CO SLED detected in the very central region of  APM 08279+5255 requires an XDR component dominating the high $J$ CO emission lines \citep{braford11}. 
We also include three $z \gtrsim 6$ quasars that are detected in at least 4 CO transitions:  J1148+5251,  J0439+1634, and J0100+2802 (\citealt{bertoldi03}; \citealt{walter03}; \citealt{beelen06}; \citealt{riechers09}; \citealt{gallerani14}; \citealt{wangf19}; \citealt{yang19}). 
Together with J2310+1855, this allows us to do a systematic study of the CO emission lines in the quasar-starburst systems at the highest redshift. 
Figure \ref{figure6} (c) shows the CO SLED of J2310+1855 compared to other high redshift quasars. APM 08279+5255 exhibits the most extreme CO SLED that is detected in the nuclear 550 pc size scale, and it represents a highly excited nuclear CO SLED exposed to the intense X-rays from the quasar. 
The limited CO detections in J0439+1634 and Cloverleaf suggests that maybe the CO SLED peaks at $J= 8 \rm \ \sim \ 10$ similar to J2310+1855. As for J1148+5251 and J0100+2802, we are not able to determine the CO SLED peak. This is because for J0100+2802, there is no CO observation at $J= 8, 9$ and $J\textgreater 11$, while the CO SLED of J1148+5251 is observed at $J\leq 7$ and $J=17$.
%The CO SLED detected in J2310+1855 is comparable to the Cloverleaf, and exhibit as high excitation as the other $z \gtrsim 6$ quasars,  but not as extreme as APM 08279+5255. 
The comparison of CO SLEDs between J2310+1855 and other high-$z$ quasars suggest that a highly excited molecular gas component is common in the nuclear region of the quasar hosts.
However, as described above, the shape of the CO SLED of these systems are different from object to object. 
Due to the lensing effect, the CO SLED of APM 08279+5255 may represent molecular gas on $\textless$1 kpc scale.  The CO SLED of J2310+1855 is not as extreme as APM 08279+5255, and more comparable to the Cloverleaf.  When compared to the two $z \textgreater 6$ quasars, J0100+2802 and J0439+1634, that have available CO data at $J= 8 \rm \ \sim \ 11$ , J2310 +1855 is more single-peaked no flatten or turn over around $J=6$.  It is possible that the cold star forming component contribute more to the flux of the mid$-J$ CO in the cases of J0100+2802 and J0439+1634.

\subsection{$ H_{2}O$ and $OH^{+}$ emission}
%The $\rm H_{2}O$ emission is found to be bright in infrared luminous galaxies, and it can be even as luminous as CO in local and high redshift ULIRGs and dusty star forming galaxies (\citealt{yang13,yang16}; \citealt{van11}; \citealt{omont13}; \citealt{ jarugula19}).
Recent studies found linear relations between the infrared luminosity and the $\rm H_{2}O (2_{0,2}-1_{1,1})$ luminosity in local and high redshift infrared bright systems, suggesting the excitation of this water transition is dominated by infrared pumping (\citealt{van11}; \citealt{yang16}; \citealt{jarugula19}). 
%But we need to mention the intrinsic scatter of the $L_{\rm H_{2}O}$ and $L_{\rm IR}$ relation: 1) the $J =2_{0,2}\  \rm H_{2}O$ state is pumped by the far infrared photons at 101 $\mu m$ from the para $\rm H_{2}O$ ground state level, which is not directly sampled by $L_{\rm IR}$; 2)  the $\rm H_{2}O$ intensity changes with different physical and chemical parameters of the ISM, e.g., the gas density, kinematic temperature, and  $\rm H_{2}O$ abundances; 3)  the dust continuum opacity at 100 $\mu m$ and the water line velocity dispersion will also affect the $L_{\rm H_{2}O}$/$L_{\rm IR}$ ratio, e.g., higher $L_{\rm H_{2}O}$/$L_{\rm IR}$ ratios can be found in systems with large $\tau_{100 \mu m}$ or high velocity dispersion (\citealt{gonz14}; \citealt{yang16}).
We first compare the $\rm H_{2}O$ detection of J2310+1855 to the local and high redshift (U)LIRGs.
Figure \ref{figure7} shows the $L_{\rm H_{2}O}$  and $L_{\rm IR}$ for local and high redshift (U/Hy)LIRGs and AGNs. The black dashed line is the best fit to the local and high redshift (U/Hy)LIRGs presented in Figure 3 of \citet{yang16}.
We consider the infrared luminosity of J2310+1855 in two cases: 1) we adopt total infrared luminosity from both the quasar and the host galaxy ($L_{\rm IR(total)}$); 2) we use the infrared luminosity only from the host galaxy ($L_{\rm IR(galaxy)}$, \citealt{shao19}). 
The linear relation is fitted with (U/Hy)LIRGs (local and high $z$) and dusty star-forming galaxies, while all the four plotted AGNs including J2310+1855 are well below this relation. This is because the AGNs not only provide the power source of water emission but also contribute significantly to the infrared luminosity (\citealt{van11};\citealt{gonz10}).
% For Mrk231 and APM 08279+5255, previous studies imply that the central luminous AGNs not only provide the power source of water emission but also contributes significantly to the infrared luminosity (\citealt{van11};\citealt{gonz10}), e.g., in Mrk231, the AGN contributes to $>50 \%$ of the total infrared luminosity. 
As for the host galaxy infrared luminosity case, J2310+1855 reveals a slightly higher water luminosity given its galaxy IR luminosity compared to the linear relation. 
%It has the largest water luminosity in its IR luminosity range among all the high redshfit sources detected in $\rm H_{2}O (2_{0,2}-1_{1,1})$.
The velocity dispersion map of the $\rm H_{2} O$  shows a velocity dispersion of $\textless 100\ \rm km\ s^{-1}$, suggesting that it may not be the large velocity dispersion that contributes to the luminous water emission. Higher spatial resolution observations are required to confirm this.
We find in J2310+1855 slightly higher $L_{\rm H_{2}O}$/$L_{\rm IR}$ ratio than that for local and high redshift (U/Hy)LIRGs.
At $z\sim$ 6, only a few quasars are detected in water emission (\citealt{banados15}; \citealt{yang19}).
Similar result is found in the $z=6.52$ quasar J0439+1634, where a higher $L_{\rm H_{2}O(3_{2,1}-3_{1,2})}$/$L_{\rm IR}$ ratio is found as compared to the linear relation \citep{yang19} .

%I.e., in the typical PDR Orion bar, the $\rm H_{2} O$/CO $(6-5)$ ratio is $\sim$ 0.2 (\citealt{habart10}; \citealt{putaud19}), while in the XDR dominate chemistry, $\rm H_{2} O$/mid-$J$ CO ratio can be as large as unity, and one of the example is Mrk 231, which has a $\rm H_{2} O$/mid-$J$ CO ratio of 1.1 \citep{gonz10}. M82 0.06 \cite{kamenetzky12}. NGC 6240 0.19 \cite{meijerink13}.   

To further investigate the heating sources of molecular gas, we also study the ratio between  $\rm H_{2}O$ and CO in J2310+1855. 
Extremely luminous $\rm H_{2} O$ emission is not expected in typical PDRs.
E.g., in the Orion bar (a representative dense PDR with $n \sim 10^{5} \ \rm cm^{-3}$ illuminated by an intense FUV radiation field of $\rm G_{0} = 4 \times 10^{4}$), the $L_{\rm H_{2}O}$/$L_{CO\ (6-5)}$ ratio is 0.20 (\citealt{habart10}; \citealt{putaud19}).
Another example is the local starburst galaxy M82 that shows a ratio of 0.06 \citep{kamenetzky12}.
If the physical/chemical condition is dominated by shocks or X-rays, then the $L_{\rm H_{2}O}$/$L_{CO\ (6-5)}$ ratio can be even as high as unity.
NGC 1266 is an S0 galaxy highly excited in molecular gas, and \citet{pellegrini13} found a $L_{\rm H_{2}O}$/$L_{CO\ (6-5)}$ ratio of 0.96 that can be only explained by shocks.
Mrk 231 is a representative of the molecular gas heated by X-rays, and it has a $L_{\rm H_{2}O}$/$L_{CO\ (6-5)}$ ratio of 1.10 $\pm$ 0.17 \citep{gonz10}.
The quasar J2310+1855 exhibits a $L_{\rm H_{2}O}$/$L_{CO\ (6-5)}$ ratio of 0.97 $\pm$ 0.09, which is comparable to Mrk 231 and NGC 1266 but is much higher than Orion bar and M82. These suggest that the molecular gas heating is not likely dominated by PDR.
In addition, we also consider the high redshit lensed SMGs for comparison (\citealt{omont13}; \citealt{yang13, yang16, yang17}). These lensed SMGs are all starbursts that are exposed to intense radiation field and shocks, and some have extremely luminous water emission lines that can not be purely explained by PDR. For the majority of the SMGs, the $L_{\rm H_{2}O}$/$L_{CO\ (6-5)}$ ratio is found to be less than 0.8, while only quite a few objects have ratios of $\sim$ 1 that is unlikely PDR.
The even higher $L_{\rm H_{2}O}$/$L_{CO\ (6-5)}$ ratio of J2310+1855 compared to typical high redshift lensed SMGs might suggest additional gas heating by the central luminous quasar.
We also compare the $\rm H_{2} O$/CO ratio of the quasar with other $z\sim 6$ quasars that are detected in water emission.
J0439+1634 was detected in the $\rm H_{2}O(3_{2,1}-3_{1,2})$ emission, and suggest a $L_{\rm H_{2}O(3_{2,1}-3_{1,2})}/L_{CO(6-5)}$ ratio of 1.23 $\pm$ 0.22. Adopting a mean $\rm H_{2}O(3_{2,1}-3_{1,2})$/$\rm H_{2}O(2_{0,2}-1_{1,1})$ of 1.4 estimated from SMGs in \citet{yang16}, we estimate the $L_{\rm H_{2}O}$/$L_{CO\ (6-5)}$ ratio in J0439+1634 of 0.86 $\pm$ 0.17, which is comparable to the value of J2310+1855.

We tentatively detected the $\rm OH^{+}$ line at a S/N of 4. The spectra of the $\rm OH^{+}$ line shows a P Cygni  like profile, with a possible absorption in the line center frequency. This might hint on possible outflows/inflows in this source, although further high sensitivity observations are needed to confirm this. 
The ratios between $\rm OH^{+}$, $\rm {H_2 O}^{+}$ and $\rm {H_3 O}^{+}$ reactive molecular ions are ideal tracers of the ionization rate and serve as the ionization source diagnostics. In J2310+1855, we are not able to constrain the ionization rate with the $\rm OH^{+}$ detection alone. On the other hand, $\rm OH^{+}$ and $\rm H_{2} O$ are all oxygen-hydrogen species, and $\rm OH^{+}$ can be formed by photodissociation of $\rm H_{2} O$, accordingly we are expecting higher $L_{\rm OH^{+}}$/$L_{\rm H_{2}O}$ ratio in the presence of cosmic rays or X-rays.
Mrk 231 is one of the best studied AGNs that has been detected in a series of $\rm OH^{+}$, $\rm {H_2 O}^{+}$ and $\rm {H_3 O}^{+}$ emission lines. 
\citet{gonz18} detected very bright $\rm OH^{+}$ emission with $L_{\rm OH^{+}}$/$L_{\rm H_{2}O}$ ratio of 0.37 $\pm$ 0.13. The ionization rate derived by making use of all the molecular ions detected are very high, and even can not be explained by its observed X-ray flux and requires ionization by cosmic rays (i.e., the ionization rate produced by X-ray photons is $\sim 1/10$ the value required).
The quasar J2310+1855 exhibit a $L_{\rm OH^{+}}$/$L_{\rm H_{2}O}$ ratio of 0.20 $\pm$ 0.15 comparable with Mrk 231. 
The observed  X-ray luminosity in J2310+1855 is $\sim 170 \times$ that observed in Mrk 231 \citep{vito19}.
If we simply assume similar ionization rate in J2310+1855 as that of Mrk 231 (as the $L_{\rm OH^{+}}$/$L_{\rm H_{2}O}$ ratios are comparable in these two AGNs), the X-ray photons from J2310+1855 are more than adequate to explain the observed $\rm OH^{+}$ emission.

\subsection{Molecular gas heating mechanisms} \label{mechanisms}
In the CO SLED modeling of J2310+1855, we find the mid to high $J$ ($J \geq 5$) CO emission lines are dominated by a ``warm" and ``dense" gas component with $T \gtrsim 150$ K, $log (n(\rm H_{2})/\rm cm^{-3} ) \gtrsim 5 $ and $log(N(\rm CO)/\rm cm^{-2}) \gtrsim 15.0$. %Although we are not able to rule out a ``cold" component with similar temperature as the cold dust that dominate in molecular gas mass but hardly contribute to CO fluxes at  $J \geq 5$. 
Such ``warm" and ``dense" gas component is warmer and denser than typical values found in local (U)LIRGs \citep{papado12, papado13}, and comparable to the extreme cases, e.g., the NGC 1068 circumnuclear disk (\citealt{krips11}; \citealt{viti14} ), and highly excited (U)LIRGs \citep{papado12, papado13}. 
Studies of local and high redshift systems suggest that the molecular gas can be heated through 1) the UV heating from young massive stars or AGNs, 2) mechanical heating by shocks generated from supernovae or AGN outflows, 3) cosmic ray heating from supernovae or AGNs, or 4) X-ray heating from the AGNs ( \citealt{bradford03}; \citealt{braford11}; \citealt{spinoglio12};  \citealt{meijerink13}; \citealt{rosen15}; \citealt{uzgil16}). We inspect the most possible mechanism that contributes to the high CO excitation and luminous $\rm H_{2}O$ and $\rm OH^{+}$ emission lines observed in J2310+1855.

To investigate the PDR origin, we fit the CO SLED of J2310+1855 to the PDR grid \citep{meijerink07}, and the best fitting result indicates a FUV flux of $\rm G_{0} = 1.0 \times 10^{4}$ and $n \sim 5.6 \times 10^{5} \ \rm cm^{-3}$. This suggests a higher FUV flux than that reported in \citet{carniani19} due to the lack of information from the $J \geq 10$ CO lines in their study.
Although the CO emission can be explained by a dense PDR illuminated by an intense FUV radiation field, the bright $\rm H_{2}O$ and $\rm OH^{+}$ emission lines (high  $\rm H_{2}O$/CO and $\rm OH^{+}$/$\rm H_{2}O$ ratio) hint at different physical and chemical conditions than in a PDR, suggesting rather the presence of X-rays, shocks or cosmic rays that heat and ionize the molecular gas. 
In the nuclear region, both the star formation activity and the luminous quasar are capable of influencing the physical and chemical conditions of the molecular gas.
The powerful quasar is able to participate in all the possible gas heating mechanisms through the X-ray it radiates, the shocks generated by AGN outflows and the cosmic rays it produces. 
%Future  $\rm OH$, $\rm {H_2 O}^{+}$ and $\rm {H_3 O}^{+}$ observations, possibly at higher spatial resolution are crucial to discriminate these possible heating and ionizing sources.

\section{Summary} \label{conclusions}
We report new detections of CO $(5-4)$, $(6-5)$, $(12-11)$ and $(13-12)$ with NOEMA and  CO $(8-7)$, $(9-8)$, $\rm H_{2}O$ and $\rm OH^{+}$  with ALMA in the $z = 6.003$ quasar J2310+1855. This is the most complete SLED ever obtained for a $z \geq 6$ quasar. 
We spatially resolved the CO $(8-7)$, $(9-8)$ and $\rm H_{2}O$ lines at similar source sizes of $\sim 2\ \rm kpc$ in FWHM, which are slightly more compact than the [\ion{C}{2}]$\ 158 \mu m$ and CO $(2-1)$ emission lines. These suggest that the high $J$ CO lines and  the $\rm H_{2}O$ line are probing the nuclear dense molecular regions closer to the quasar. 
We analyze the physical conditions of the molecular gas through CO SLED modeling and we also compare the CO emission lines from the quasar with local and high redshift starburst samples and some representative local and high redshift AGNs.
The main results are summarized below.

$\bullet$ The CO SLED of J2310+1855 at $J \geq 5$ is dominated by a ``warm" and ``dense" gas component in the parameter range of $T_{\rm kin} \approx 167^{+153}_{-56} \ \rm K$, $log (n(\rm H_{2}) / \rm cm^{-3})\approx 5.11^{+1.83}_{-0.58}$ and $log(N(\rm CO)/\rm \ cm^{-2}) \approx 17.28^{+0.33}_{-0.42} $ (in the one-component model) or $T_{\rm kin} \approx 306^{+263}_{-149}\  \rm K$, $log (n(\rm H_{2}) / \rm cm^{-3}) \approx 5.22^{+1.04}_{-0.49}$ and $log(N(\rm CO)/\rm \ cm^{-2}) \approx 15.29^{+1.34}_{-1.17}$ (in the two-component model). We are not able to rule out a ``cold" component that dominates the molecular gas mass but barely contributes to the $J \geq 5$ CO fluxes. 

$\bullet$ The CO SLED of J2310+1855 shows higher excitation compared to local/high redshift starburst samples and local AGNs. Such high CO excitation is also found in other $z \gtrsim 6$ quasars (e.g., J1148+5251, J0100+2802, J0439+1634), and lensed high redshift quasars (e.g., APM 08279+5255, the Cloverleaf).

$\bullet$ The $L_{\rm H_{2}O}/L_{\rm IR(galaxy)}$ ratio in this quasar is higher than local and high $z$ (U/Hy)LIRGs. The luminous detections of $\rm H_{2}O$ and $\rm OH^{+}$ (high  $\rm H_{2}O$/CO and $\rm OH^{+}$/$\rm H_{2}O$ ratios) are suggesting other heating and ionization sources (e.g., cosmic rays, shocks and X-rays) in addition to PDR. In the nuclear region, the luminous quasar and the starburst activity are able to impact on the molecular gas through all these possible mechanisms.

Complete measurements of the CO SLED  of the quasar hosts at $z \geq 6$ are of great importance for our understanding of the physical conditions and the heating mechanisms of the molecular gas in the complex environment with both AGN and nuclear starburst activities. 
It is also essential for the higher resolution observations to map the distributions and kinematics of the highly excited molecular gas around the AGN.

\newpage 
\begin{figure*}
\centering 
\includegraphics[width=0.9\textwidth]{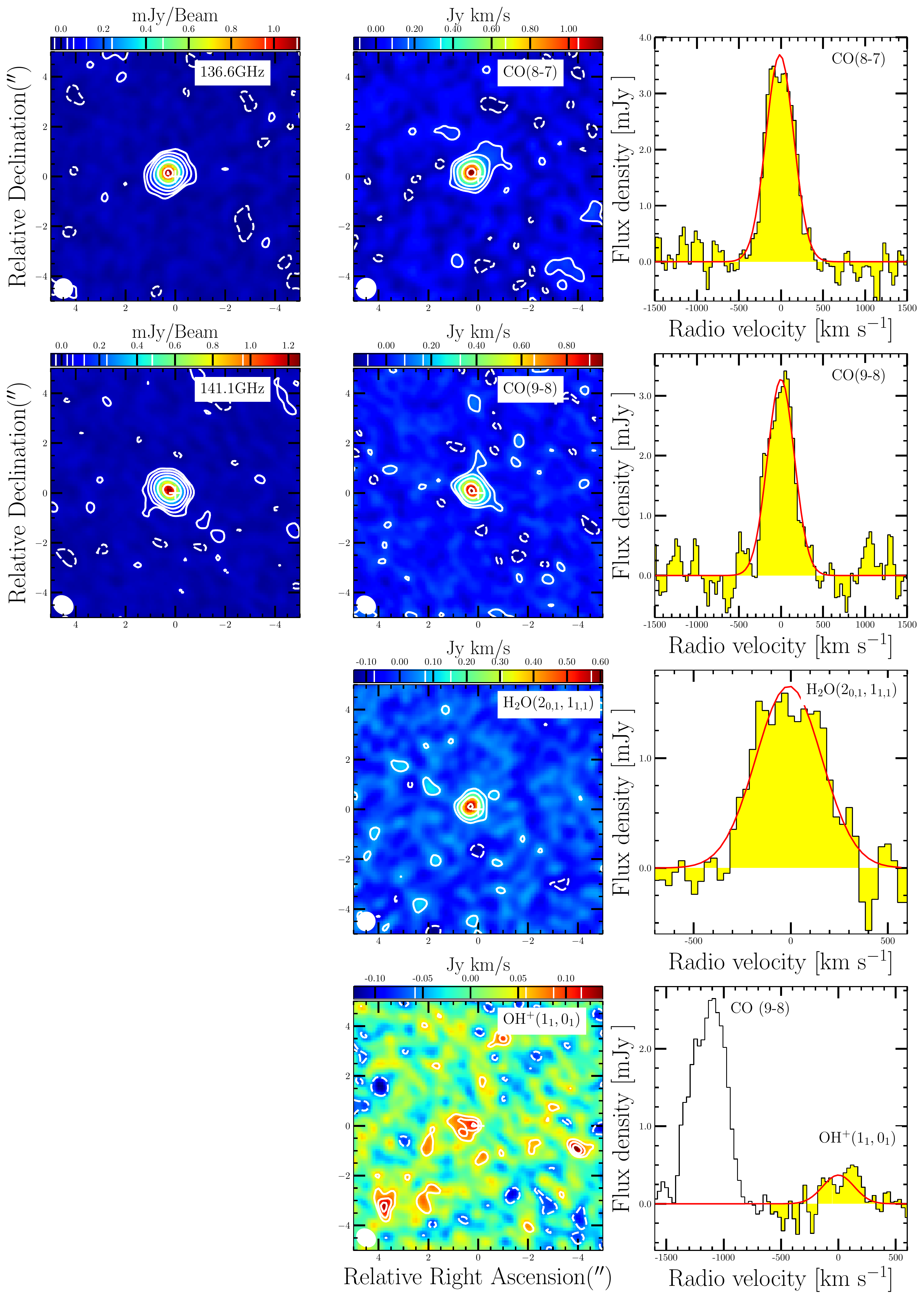}
\caption{(Caption continued on the next page.)}% First caption
\end{figure*}
\addtocounter{figure}{-1}
\begin{figure*}
%\centering
%\includegraphics[width=0.8\textwidth]{figure_2.pdf}
\caption{ Continuum, line intensity maps, and spectrum ({\bf from left to right}) of CO $(8-7)$, $(9-8)$, $\rm H_{2}O (2_{0,2}-1_{1,1})$ and $\rm OH^{+} (1_{1}-0_{1})$ ({\bf from top to bottom}) observed by ALMA.  
The white cross represents the Gaia position of the quasar \citep{shao19}. The size of the white cross demonstrate the astrometric uncertainty of the quasar position. The filled white ellipse on the lower left shows the FWHM of the beam.
{\bf First column:} Continuum maps. The white contours denote [-2, 2, 4, 8, 16, 32, 64, 74]$\times \sigma$ (1$\sigma$ = 15 $\mu$Jy beam$^{-1}$) at 136.6 GHz and [-2, 2, 4, 8, 16, 32, 64]$\times \sigma$ (1$\sigma$ = 15 $\mu$Jy beam$^{-1}$) at 141.1 GHz.
{\bf Second column:} Spectra line intensity maps. The white contours denote  [-2, 2, 4, 8, 16, 25]$\times \sigma$ (1$\sigma$ = 0.042 mJy beam$^{-1} \cdot \rm km\ s^{-1}$), [-2, 2, 4, 8, 16, 22]$\times \sigma$ (1$\sigma$ = 0.041 mJy beam$^{-1} \cdot \rm km\ s^{-1}$), [-2, 2, 4, 8, 15]$\times \sigma$ (1$\sigma$ = 0.038 mJy beam$^{-1} \cdot \rm km\ s^{-1}$) and [-2, 2, 3, 4]$\times \sigma$ (1$\sigma$ = 0.029 mJy beam$^{-1} \cdot \rm km\ s^{-1}$) for CO $(8-7)$, $(9-8)$, $\rm H_{2}O (2_{0,2}-1_{1,1})$ and $\rm OH^{+} (1_{1}-0_{1})$ respectively.
{\bf Third column:} Yellow histogram represents the spectra extracted from the 2 $\sigma$ contour on the intensity map for spatially resolved CO $(8-7)$, $(9-8)$, $\rm H_{2}O (2_{0,2}-1_{1,1})$ lines, and from the peak pixel for spatially unresolved $\rm OH^{+} (1_{1}-0_{1})$ emission. The red solid line is a single Gaussian profile fit to the spectral line. The Gaussian fit to the $\rm OH^{+} (1_{1}-0_{1})$ spectral line should be taken with caution since there are possibly both emission and absorption features, and the current sensitivity is not enough to confirm these features.
The spatial offsets between the continuum and spectra line emissions are within the uncertainty of the quasar position.} 
\label{figure1}
\end{figure*}

\begin{figure*}
\centering 
\includegraphics[width=0.9\textwidth]{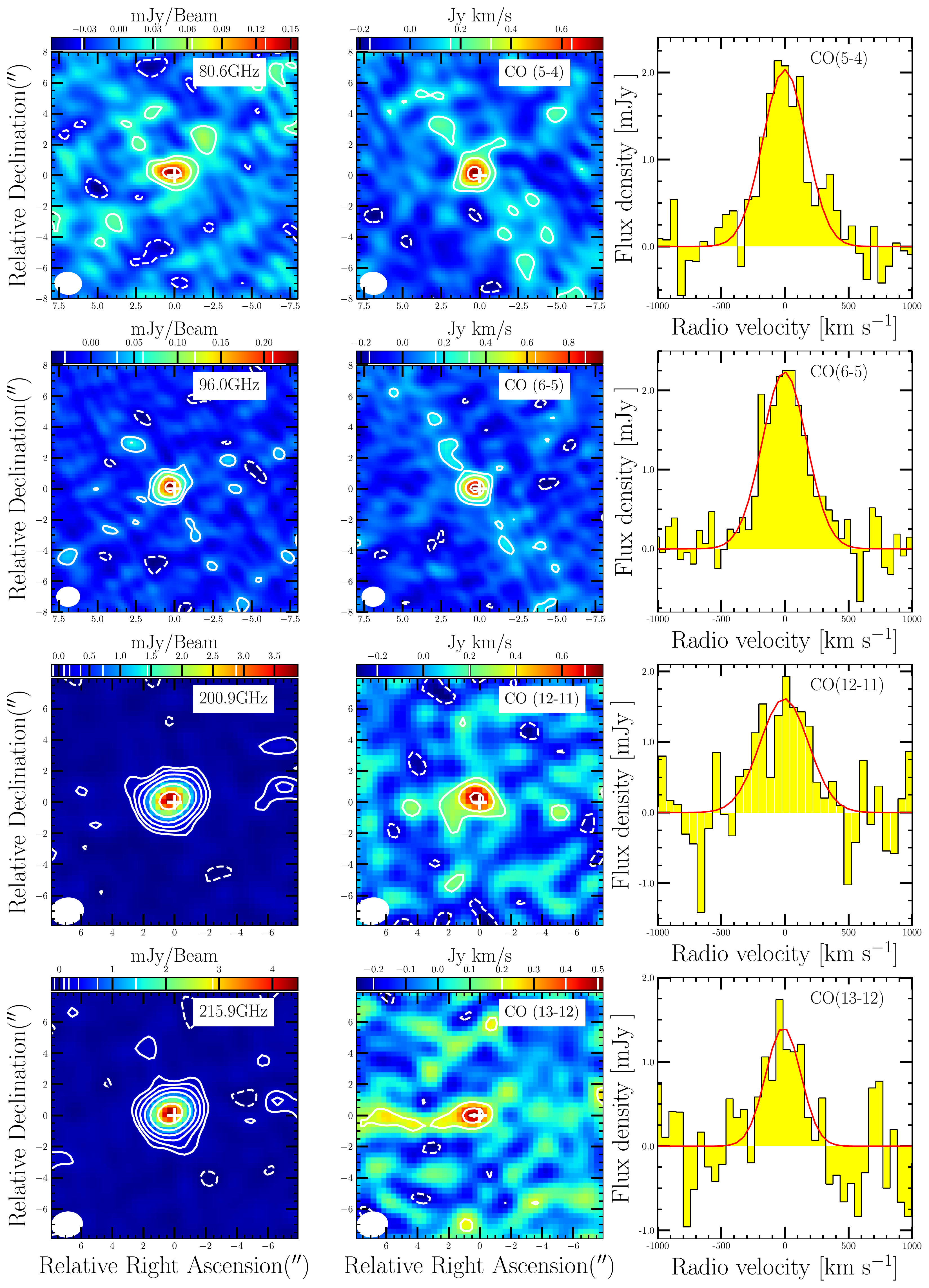}
\caption{(Caption continued on the next page.)}% First caption
\end{figure*}
\addtocounter{figure}{-1}
\begin{figure*}
%\centering
%\includegraphics[width=0.8\textwidth]{figure_1.pdf}
 \caption{Continuum,  line intensity maps, and spectrum ({\bf from left to right}) of CO $(5-4)$, $(6-5)$, $(12-11)$ and $(13-12)$ 
({\bf from top to bottom}) observed by NOEMA. 
The white cross on continuum and line intensity map represents the Gaia position of the quasar \citep{shao19}. The size of the white cross demonstrates the quasar location astrometric uncertainty. The filled white ellipse on the lower left shows the FWHM of the beam.
For the continuum maps ({\bf First column}), contours denote [-2, 2, 4, 8]$\times \sigma$ (1$\sigma$ = 16 $\mu$Jy beam$^{-1}$) at 80.6 GHz, [-2, 2, 4, 8, 14]$\times \sigma$ (1$\sigma$ = 15 $\mu$Jy beam$^{-1}$) at 96.0 GHz,  [-2, 2, 4, 8, 16, 32, 64]$\times \sigma$ (1$\sigma$ = 45 $\mu$Jy beam$^{-1}$) at 200.9 GHz and [-2, 2, 4, 8, 16, 32, 64]$\times \sigma$ (1$\sigma$ = 45 $\mu$Jy beam$^{-1}$) at 215.9 GHz. 
For the spectra line intensity maps ({\bf Second column}), contours denote [-2, 2, 4, 8, 10]$\times \sigma$ (1$\sigma$ = 0.08 mJy beam$^{-1} \cdot \rm km\ s^{-1}$), [-2, 2, 4, 8, 11]$\times \sigma$ (1$\sigma$ = 0.08 mJy beam$^{-1} \cdot \rm km\ s^{-1}$), [-2, 2, 4, 7]$\times \sigma$ (1$\sigma$ = 0.10 mJy beam$^{-1} \cdot \rm km\ s^{-1}$) and [-2, 2, 4, 5]$\times \sigma$ (1$\sigma$ = 0.10 mJy beam$^{-1} \cdot \rm km\ s^{-1}$) for  CO $(5-4)$, $(6-5)$, $(12-11)$ and $(13-12)$ respectively.
{\bf Third column}: Yellow histogram represents spectrum extracted from the peak pixel (all spectra lines are spatially unresolved), and the red solid line is a single Gaussian profile fit to the spectral line. The peak positions of the continuum and spectra line emissions are within uncertainty of the quasar position.}
\label{figure2}
\end{figure*}

\begin{figure*}
\includegraphics[width=1.0\textwidth]{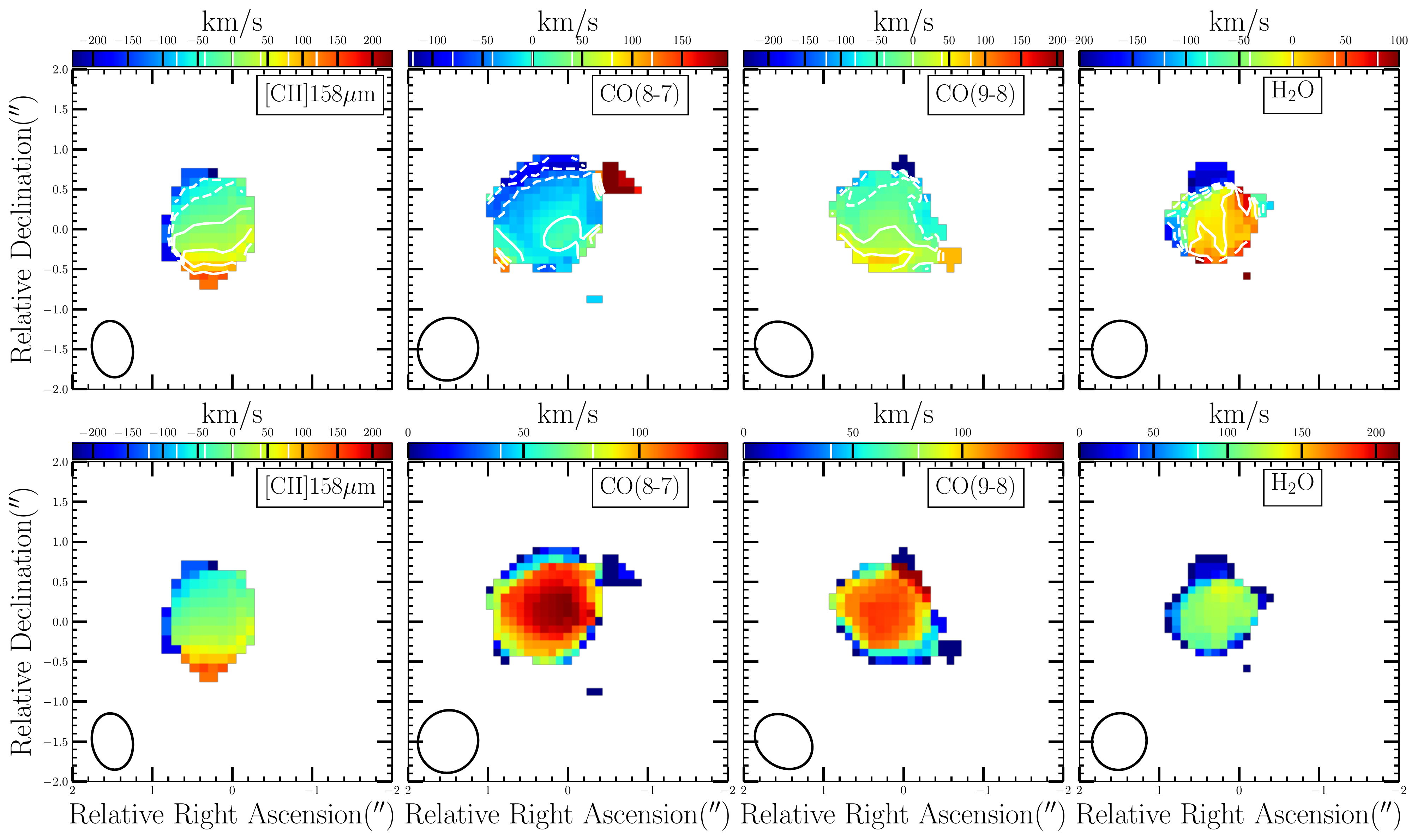}
\caption{Velocity ({\bf first row}) and velocity dispersion ({\bf second row}) of [\ion{C}{2}]$\rm\ 158 \mu m$, CO $(8-7)$, $(9-8)$ and $\rm H_{2}O (2_{0,2}-1_{1,1})$ (Left to right). The black ellipse on the left shows the FWHM of the beam.
We generate the velocity and velocity dispersion maps with the pixels of $\geq 3.0 \sigma$ values in the line emitting channels, and the zero velocity corresponds to the [\ion{C}{2}]$\rm\ 158 \mu m$ redshift of 6.0031 \citep{wang13}.
White contours on the velocity maps are [-4, -2, 0, 2, 4, 6] $\times 20\  \rm km\ s^{-1}$ for  [\ion{C}{2}]$\rm\ 158 \mu m$,  [-6, -4, -2, 0, 2, 4, 6] $\times 20\  \rm km\ s^{-1}$ for CO $(8-7)$, [-6, -4, -2, 0, 2, 4] $\times 20\  \rm km\ s^{-1}$ for CO $(9-8)$, and [-6, -4, -2, 0, 2, 4] $\times 20\  \rm km\ s^{-1}$ for $\rm H_{2}O (2_{0,2}-1_{1,1})$. 
We find in the velocity maps that CO $(8-7)$, $(9-8)$ and $\rm H_{2}O (2_{0,2}-1_{1,1})$ generally follow the velocity gradient observed from  [\ion{C}{2}]$\rm\ 158 \mu m$  \citep{wang13}  from northeast to southwest.}
\label{figure3}
\end{figure*}

\begin{figure*}
\includegraphics[width=0.5\textwidth]{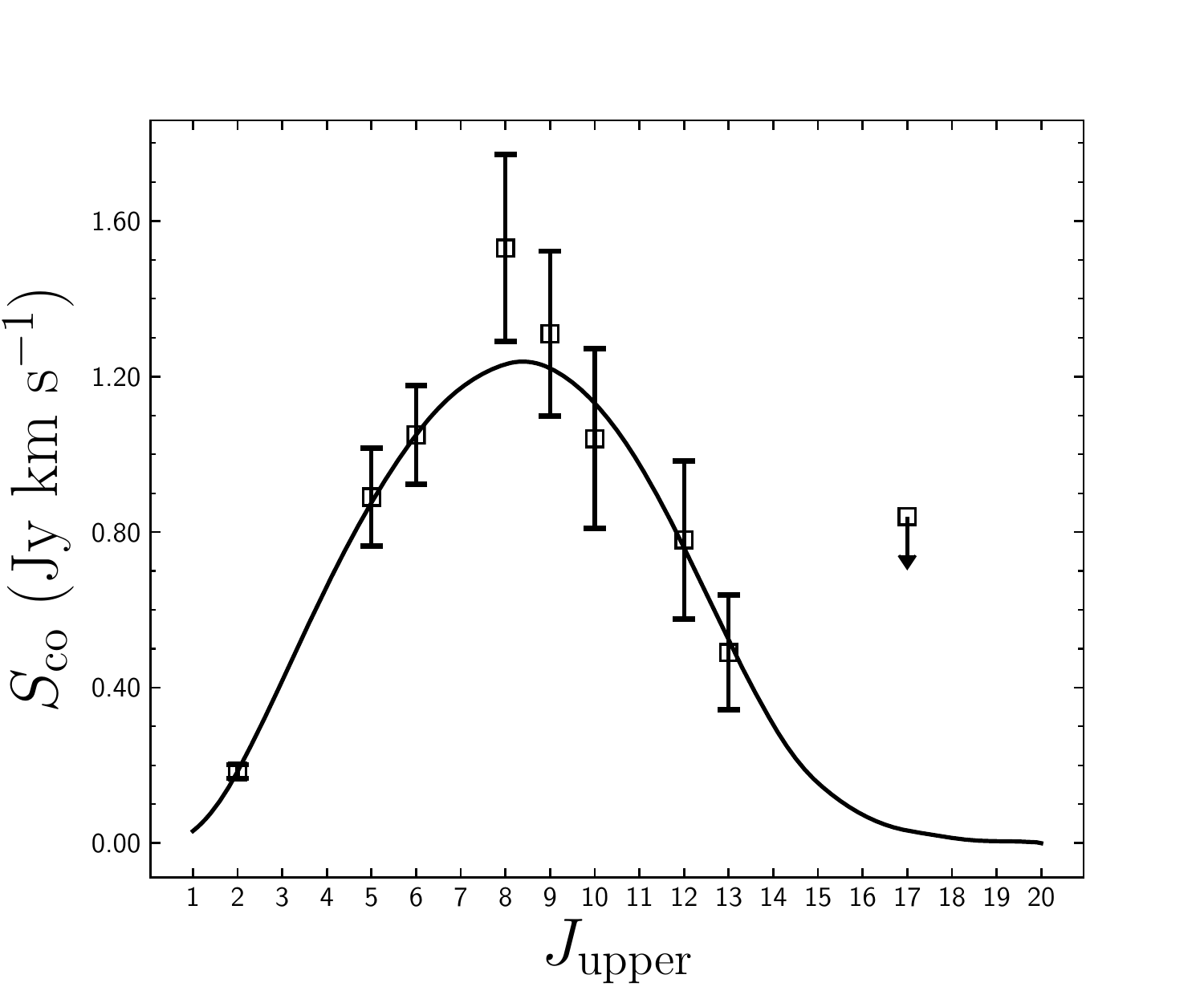}
\includegraphics[width=0.5\textwidth]{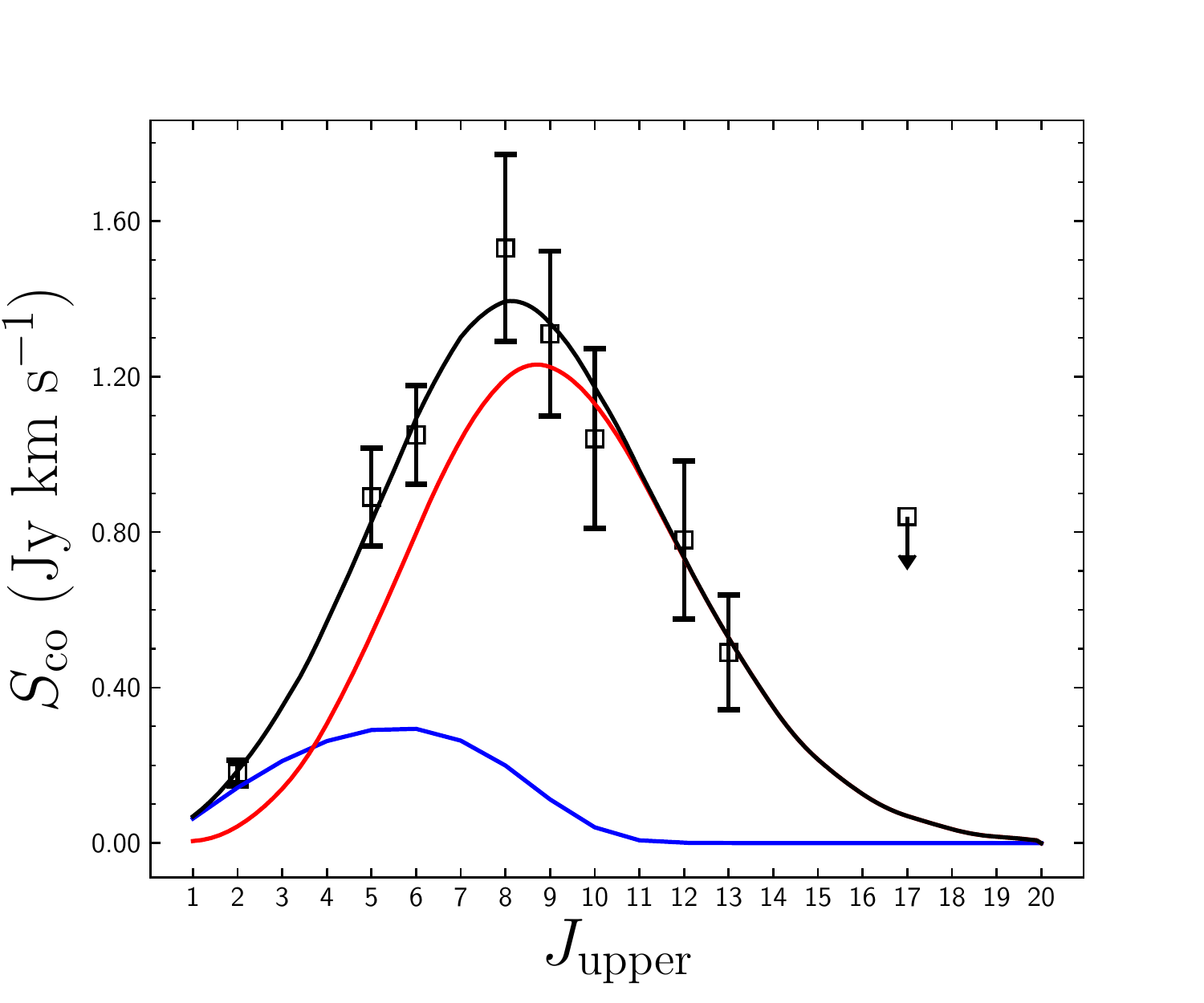}
\caption{CO SLED fitted with one ({\bf{left}}) or two component ({\bf{right}}) models. Black squares are the CO fluxes of J2310+1855 with the calibration uncertainties included. The CO $(5-4)$, $(6-5)$, $(8-7)$, $(9-8)$, $(12-11)$ and $(13-12)$ data are from this work. The CO $(2-1)$ data is taken from \citet{shao19}, the CO $(10-9)$ data will be presented in Riechers et al. (in prep), and the upper limit of the CO $(17-16)$ line is from \citet{carniani19}. Details are presented in Table \ref{table2}. 
{\bf{Left}:} Black solid line represents the best one-component fitting result with the minimum $\chi^{2}$ ($T_{\rm kin} = 228 \ \rm K$, $log (n(\rm H_{2})/\rm cm^{-3}$ )=4.75 and $log(N(\rm CO)/\rm cm^{-2}) =17.5$). 
{\bf{Right}:} Least squares fitting result with two-component model. Blue solid line represents the J1148+5251 model with $T_{\rm kin} = 50 \ \rm K$, $log (n(\rm H_{2})/\rm cm^{-3} )=4.20$ and $log(N(\rm CO)/\rm \ cm^{-2}) =18.0 $ \citep{riechers09}. 
Red solid line is the ``warm" component with $T_{\rm kin} = 306 \ \rm K$, $log (n(\rm H_{2})/\rm cm^{-3} )=5.25$ and $log(N(\rm CO)/\rm cm^{-2}) =15.5 $.
}
\label{figure4}
\end{figure*}

\begin{figure*}
\includegraphics[width=0.5\textwidth]{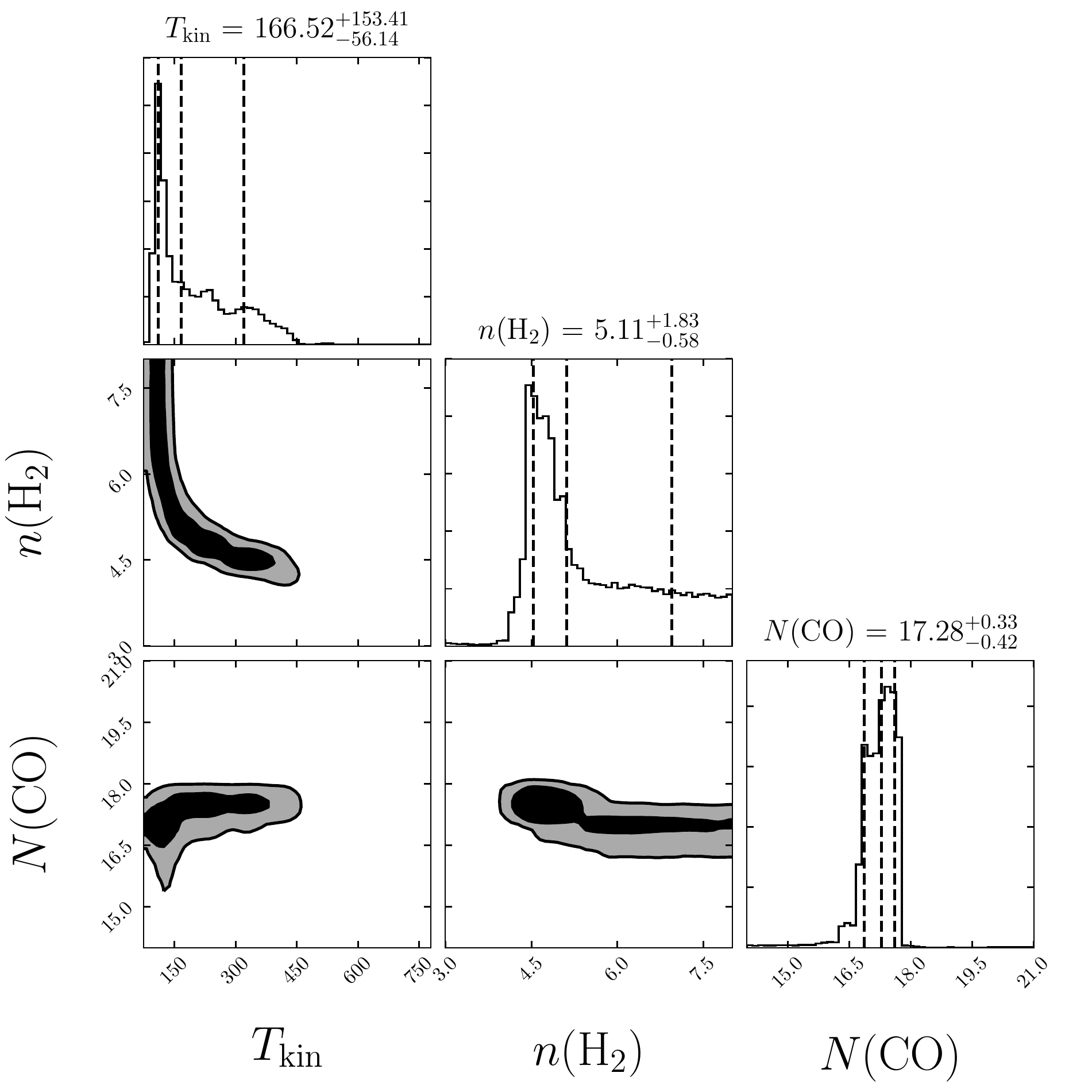}
\includegraphics[width=0.5\textwidth]{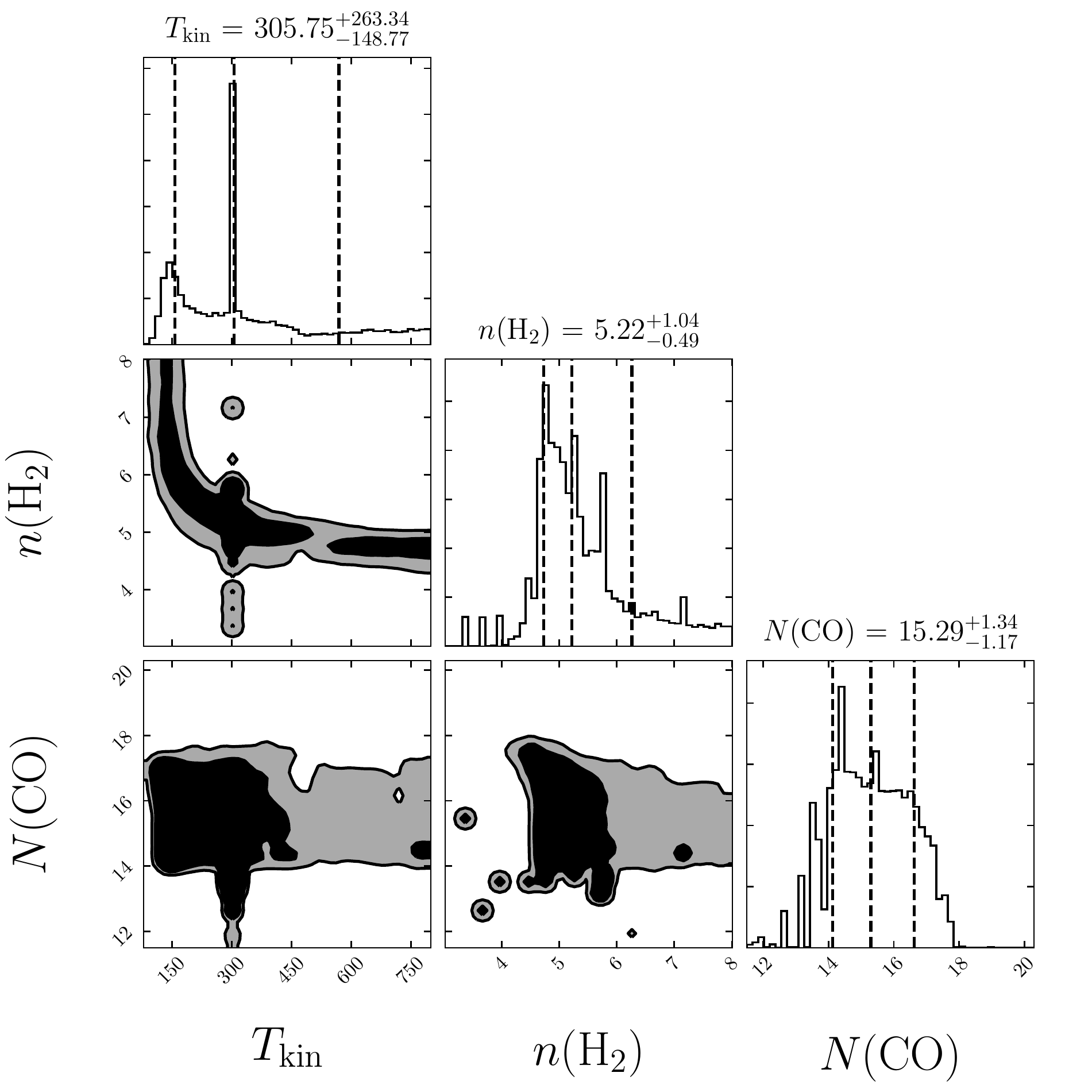}
\caption{ Posterior probability distributions of the three parameters $T_{\rm kin}$  (K), $log (n(\rm H_{2})/\rm cm^{-3})$ and $log(N(\rm CO)/\rm cm^{-2})$. The plotted contours show 95 $\%$ and 68 $\%$ confidence intervals. {\bf{Left:}} Posterior probability distribution of the parameters for the one component model. The resulting MCMC result is $T_{\rm kin} \approx 167^{+153}_{-56} \ \rm K$, $log (n(\rm H_{2}) / \rm cm^{-3})\approx 5.11^{+1.83}_{-0.58}$ and $log(N(\rm CO)/\rm \ cm^{-2}) \approx 17.28^{+0.33}_{-0.42} $. We note that the median and uncertainties here are calculated based on the 16th, 50th, and 84th percentiles of the samples in the marginalized distributions.  
{\bf Right:} Posterior probability distribution of the ``warm" component in the  two-component model fit to the data. The fitting result suggests the ``warm" component with $T_{\rm kin} \approx 306^{+263}_{-149}\  \rm K$, $log (n(\rm H_{2}) / \rm cm^{-3}) \approx 5.22^{+1.04}_{-0.49}$ and $log(N(\rm CO)/\rm \ cm^{-2}) \approx 15.29^{+1.34}_{-1.17}$.} 
\label{figure5}
\end{figure*}

\begin{figure*}
\includegraphics[width=1.0\textwidth]{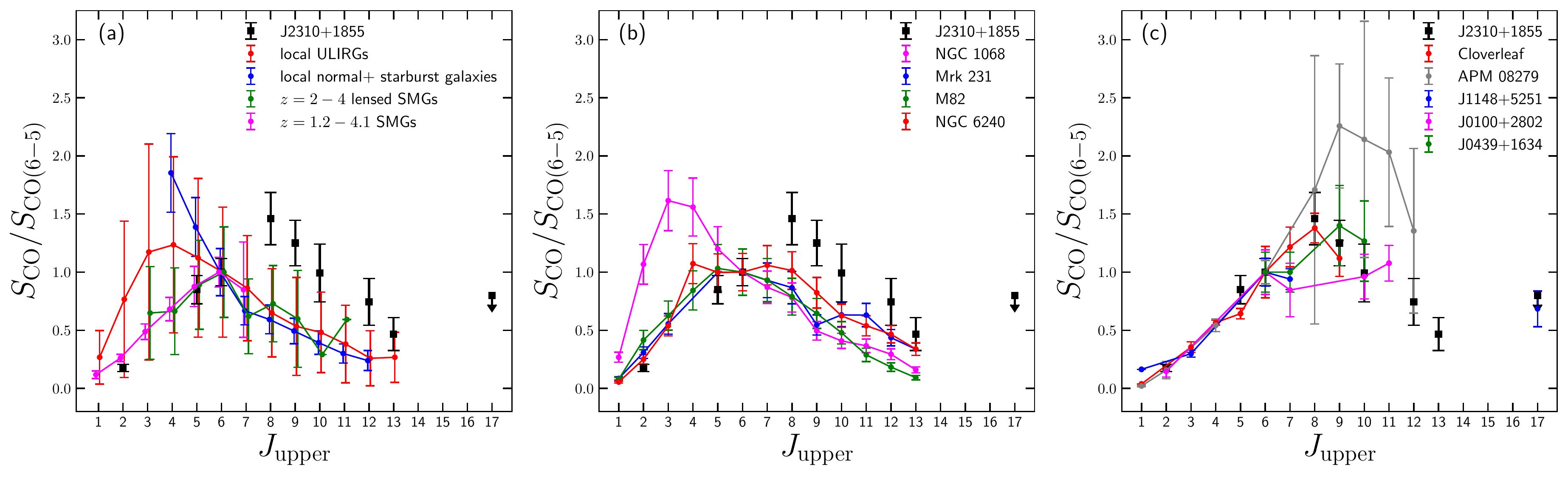}
\caption{CO SLED normalized to CO $(6-5)$ in normalized unit of Jy $\rm km\ s^{-1}$. 
{\bf{Left column:}} J2310+1855 (black squares) in comparison with the mean of four galaxy samples: the local (U)LIRGs \citep{rosen15} (red); local normal + starburst galaxies \citep{liu15} (blue); $z \sim 1.2-4.1$ SMGs \citep{bothwell13} (magenta); strongly lensed SMGs at $z \sim 2-4$ \citep{yang17} (green). 
{\bf{Middle column:}} J2310+1855 (black squares) in comparison with local starburst systems and AGNs.  M82 (\citealt{panuzzo10}; \citealt{weiss05}) (green) is a representative example of the local starburst galaxy. The local representative AGNs are NGC 1068 \citep{spinoglio12} (magenta), Mrk231 \citep{van10} (blue) and NGC 6240 \citep{rosen15}(red).
{\bf{Right column:}} J2310+1855 (black squares) in comparison with high redshift quasars. The plotted quasars are APM 08279+5255 (\citealt{braford11}; \citealt{weiss07}; \citealt{riechers09}) (grey), Cloverleaf (\citealt{bradford09}; \citealt{uzgil16}) (red), J1148+5251 (\citealt{bertoldi03}; \citealt{walter03}; \citealt{beelen06}; \citealt{riechers09}; \citealt{gallerani14}) (blue), J0439+1634 \citep{yang19} (green) and J0100+2802 \citep{wangf19} (magenta). 
} 
\label{figure6}
\end{figure*}

\newpage
\begin{figure*}
\includegraphics[width=0.5\textwidth]{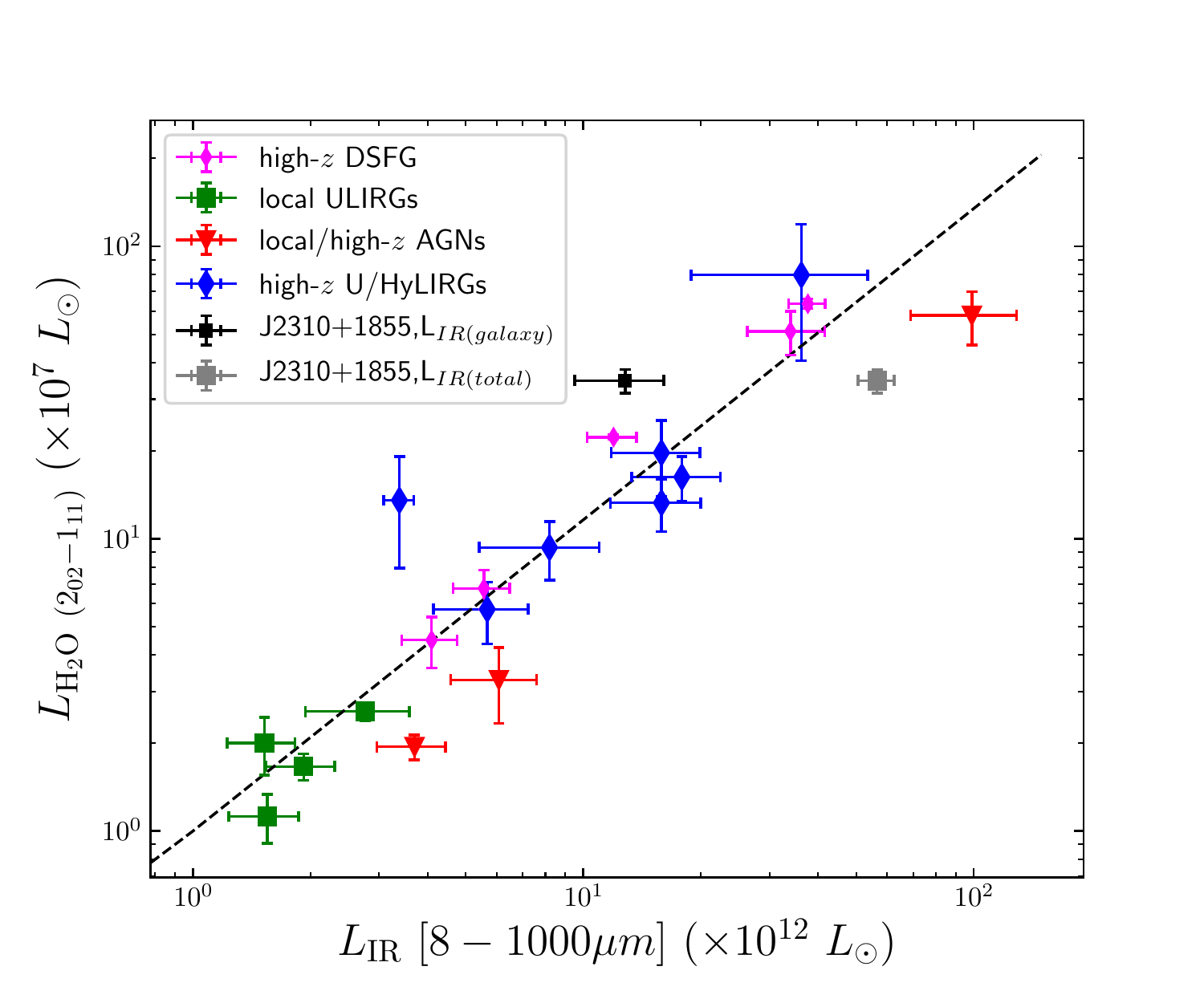}
\includegraphics[width=0.5\textwidth]{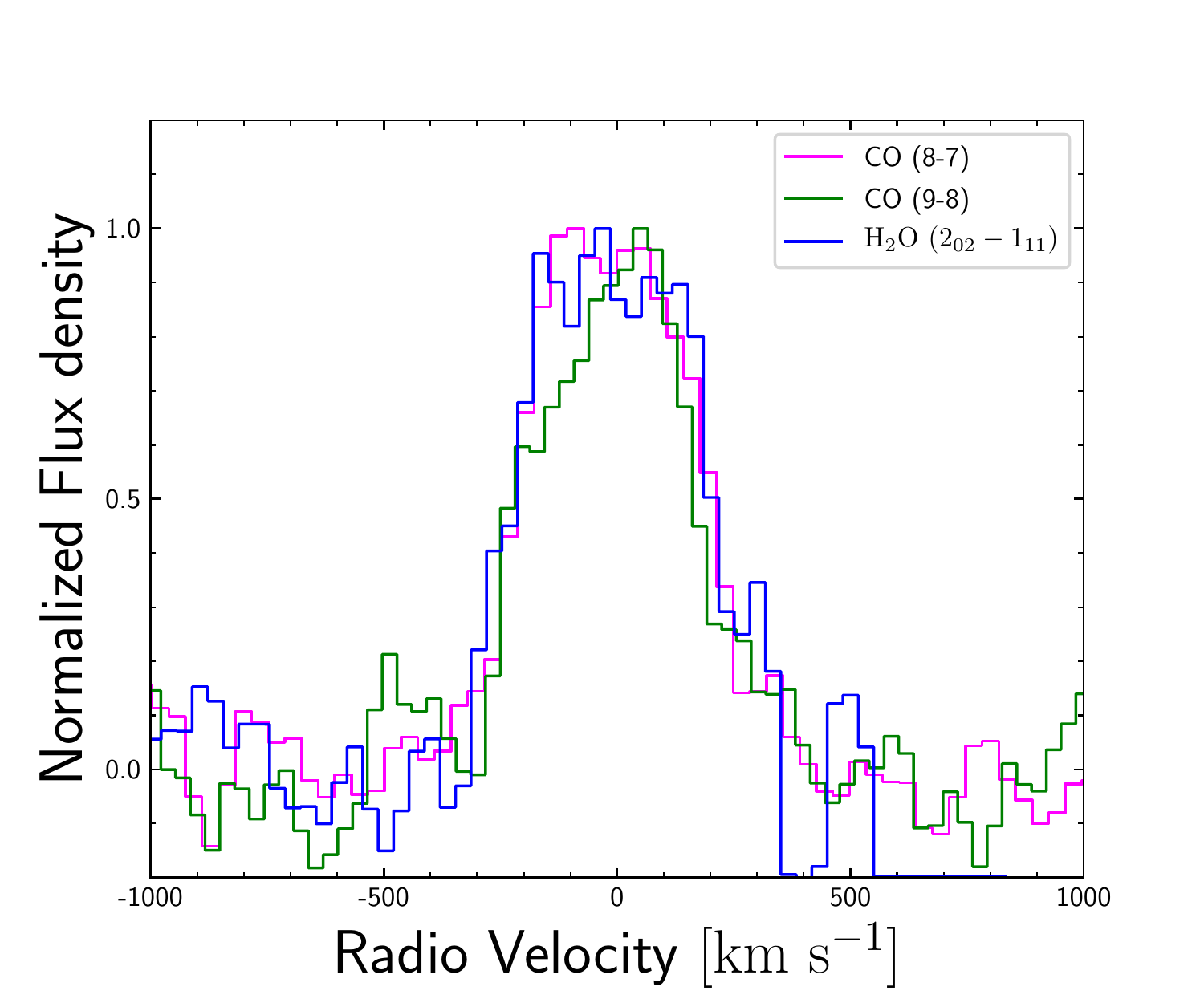}
\caption{{\bf Left}: $L_{\rm H_{2}O (2_{0,2}-1_{1,1})}\ vs \ L_{\rm IR}$ (from Fig.3 in \citealt{yang16}). 
The green squares are the local ULIRGs from \citet{yang13}, the blue diamonds are the high redshift U/HyLIRGs from \citet{yang16}, \citet{omont13}, and \citet{van11}, the magenta diamonds are the high redshift (dusty) star forming galaxies from \citet{jarugula19} and \citet{apostolovski19}, and the red down-triangles represent local and high redshift AGNs, namely Mrk 231 \citep{gonz10}, SDP81 \citep{yang16} and APM 08279+5255 \citep{braford11} from left to right. Note that all the luminosities plotted are intrinsic luminosities that have been corrected for lensing.
The grey and black squares mark two cases of J2310+1855: the former (grey square) is the water line to total infrared luminosity ratio with $L_{\rm IR(total)} $ = (5.7 $\pm$ 0.6) $\times \ 10^{13} \ L_{\sun}$ (with contributions from both quasar and host galaxy), and the latter (black square) shows the water line to galaxy infrared luminosity ratio with $L_{\rm IR(galaxy)}$ = (1.4 $\pm$ 0.3) $\times \ 10^{13} \ L_{\sun}$ (which is purely from the host galaxy). The black dashed line represents the best fit to the local $\&$ high $z$ U/HyLIRGs (green squares and blue diamonds) with $L_{\rm H_{2}O (2_{0,2}-1_{1,1})}\sim \ L_{\rm IR}^{1.06}$, see \citep{yang16}.
%The black solid line represents the best fit to the local ULIRGs, high $z$ U/HyLIRGs and dusty star forming galaxies (green squares, blue diamonds and magenta diamonds) with $L_{\rm H_{2}O (2_{0,2}-1_{1,1})}\sim \ L_{\rm IR}^{1.18}$. 
Note that the definition of infrared luminosity is $8-1000\ \mu m$. {\bf Right}: Spectrum of CO $(8-7)$ (magenta), $(9-8)$ (green) and $\rm H_{2}O (2_{0,2}-1_{1,1})$ (blue) normalized to the peak flux densities.}
\label{figure7}
\end{figure*}

\newpage
\begin{deluxetable*}{lccccccc}
%\tabletypesize{\scriptsize}
%\tablenum{1}
\tablecaption{Observational Details}
\tablecolumns{8}
\tablewidth{0pc}
%\tablecaption{Continuum properties}
\tablehead{
\colhead{Line ID}    & \colhead{$\nu _{\rm obs}$}  &\colhead{$\delta \nu$($\delta$v)}& \colhead{Band}&\colhead{$T_{\rm source}$}   &  \colhead{$T_{\rm tot}$} & rms \\ 
\colhead{}   & \colhead{GHz} & \colhead{MHz  (km $\rm s^{-1}$)} &\colhead{}   &  \colhead{mins} & \colhead{mins}& \colhead{$\rm mJy\ beam^{-1}$}\\
\colhead{(1)}&\colhead{(2)}&\colhead{(3)}&\colhead{(4)}&\colhead{(5)}&\colhead{(6)} &\colhead{(7)} }
\startdata
CO $(5-4)$ & 82.2875& 16(58)& NOEMA 3 mm &70.8& 120&0.32\\
CO $(6-5)$ & 98.7381& 16(49)& - &-& -&-\\
CO $(8-7)$ & 131.6274&15.625(36)&ALMA Band 4&34.9&53.3 &0.17\\
$\rm H_{2}O (2_{0,2}-1_{1,1})$ & 141.0699 &15.625(36)&-&- &- &-\\
CO $(9-8)$ & 148.0648&15.625(32)&ALMA Band 4&30.3&48.5 & 0.19 \\
$\rm OH^{+} (1_{1}-0_{1})$&147.5061&15.625(32)&-&-&- & - \\
CO $(12-11)$&197.3405& 40(61)&NOEMA 1 mm & 372&600&0.54\\
CO $(13-12)$&213.7515&40(56) &-&-&-&-
\enddata    
\tablecomments{Column 1: Line ID; Column 2: Line center frequency in the observer frame; Column 3: Binned spectral resolution in frequency (velocity); Column 4-7: Observing band, on source time, total observing time and achieved sensitivity per binned channel. The lines without on source time are observed in the same frequency setup as the upper ones with on source time listed in the table.}
\label{table1}
\end{deluxetable*}

%\begin{rotate}
\begin{deluxetable*}{ccccccccc}
\tabletypesize{\tiny}
\tablecaption{Spectral line Observations}
%\tablenum{2}
\tablecolumns{9}
\tablewidth{0pc}
\tablehead{
\colhead{Line}   &  \colhead{$z_{\rm line}$}  &  \colhead{FWHM} & \colhead{S$\delta v$} &\colhead{Beam Size}   &  \colhead{Source Size}  &\colhead{Luminosity}&\colhead{Facilities}&\colhead{Ref.}\\
\colhead{}   & \colhead{}  &  \colhead{(km s$^{-1}$)} & \colhead{(Jy $\rm km\ s^{-1}$)} &\colhead{(arcsec)}   &  \colhead{(arcsec)} & \colhead{($10^{9}\ L_{\sun}$)}& \colhead{}&\colhead{}\\
\colhead{(1)}&\colhead{(2)}&\colhead{(3)}&\colhead{(4)}&\colhead{(5)} &\colhead{(6)}&\colhead{(7)}&\colhead{(8)} &\colhead{(9)} }
\startdata
CO $(2-1)$&6.0029  $\pm$  0.0005&484 $\pm$ 48&0.18 $\pm$ 0.02&0.61 $\times$ 0.59&(0.60 $\pm$ 0.18) $\times$ (0.40 $\pm$ 0.21)&0.021 $\pm$ 0.002&VLA&S19\\
CO $(5-4)$&6.0030 $\pm$ 0.0004&409 $\pm$ 44& 0.89 $\pm$ 0.09 &1.67 $\times$ 1.37& - &  0.254 $\pm$ 0.026 & NOEMA&L19  \\
CO $(6-5)$&6.0025 $\pm$ 0.0007&456 $\pm$ 64&1.52 $\pm$ 0.13&5.4 $\times$ 3.9&-&  0.520 $\pm$ 0.045  &PdBI &W13\\
CO $(6-5)$&6.0028 $\pm$ 0.0003&361 $\pm$ 9&1.12 $\pm$ 0.06&0.6 $\times$ 0.4& (0.33 $\pm$ 0.06) $\times$ (0.20 $\pm$ 0.04) &   0.383 $\pm$ 0.021  &ALMA  & F18\\
CO $(6-5)$&6.0030 $\pm$ 0.0003&422 $\pm$ 30& 1.05 $\pm$ 0.07 &1.42 $\times$ 1.19&(0.74 $\pm$ 0.34) $\times$ (0.46 $\pm$ 0.28)&  0.359 $\pm$ 0.024 & NOEMA&L19  \\
CO $(8-7)$&  6.0028 $\pm$ 0.0001&390 $\pm$ 15& 1.53 $\pm$ 0.05 &0.79 $\times$ 0.75&(0.46 $\pm$ 0.09) $\times$ (0.21 $\pm$ 0.10)&  0.699 $\pm$ 0.023&ALMA& L19  \\
CO $(9-8)$& 6.0031 $\pm$ 0.0002&376 $\pm$ 18   &1.31 $\pm$ 0.06   & 0.77 $\times$ 0.63&(0.41 $\pm$ 0.10) $\times$ (0.32  $\pm$ 0.11)& 0.673   $\pm$ 0.030 & ALMA&L19  \\
CO $(10-9)$& - &-   &1.04 $\pm$ 0.17   & - & & 0.594   $\pm$ 0.097 &  -&Rpr  \\
CO $(12-11)$&  6.0030 $\pm$  0.0008 & 451$\pm$ 81   &0.78 $\pm$ 0.13   & 2.08 $\times$ 1.62&-&   0.534 $\pm$ 0.089& NOEMA&L19  \\
CO $(13-12)$& 6.0028 $\pm$ 0.0007 &  324 $\pm$ 75    &0.49 $\pm$ 0.11   & 1.91 $\times$ 1.53 &-&   0.363 $\pm$ 0.082& NOEMA&L19  \\
%CO $(10-9)$&-&-   &xx   & -&-&xx&Dominiket.al. in prep\\
$\rm H_{2}O$& 6.0028 $\pm$ 0.0003 &398 $\pm$ 28 & 0.70 $\pm$ 0.05 &  0.72 $\times$  0.68&(0.39 $\pm$ 0.14) $\times$ (0.25  $\pm$ 0.21)&    0.343 $\pm$ 0.024  & ALMA&L19 \\
$\rm OH^{+}$ & - &320 $\pm$ 313 & 0.13 $\pm$ 0.10 &  0.77 $\times$ 0.63&-&0.067 $\pm$ 0.051  &ALMA&L19\\
%$\rm [NII]_{122um}$&68.4&6.0021 $\pm$ 0.0007&342 $\pm$ 71&0.50 $\pm$ 0.10&0.56 $\times$ 0.46&(0.56 $\pm$ 0.21) $\times$ (0.38 $\pm$ 0.28)&  0.609 $\pm$ 0.122 & This work  \\
%$\rm [OI]_{145um}$&73.4&6.0027 $\pm$ 0.0002&350 $\pm$ 16&0.97 $\pm$ 0.04&0.47 $\times$ 0.39&(0.47 $\pm$ 0.08) $\times$ (0.28 $\pm$ 0.11) &  0.990 $\pm$ 0.041 & This work \\
$\rm [CII]_{158um}$&6.0031 $\pm$ 0.0002&393 $\pm$ 21&8.83 $\pm$ 0.44&0.72 $\times$ 0.51&(0.55 $\pm$ 0.05) $\times$ (0.40 $\pm$ 0.07)&8.310 $\pm$ 0.414&ALMA&W13
\enddata
\tablecomments{Column 1: Line ID; Column 2 - 4: Redshift, line width in FWHM and line flux. Note that the line flux is calculated trough a single Gaussian fit to the line profile; Column 5: Beam size in FWHM; Column 6: Source size deconvolved from the beam in FWHM; Column 7:Line luminosity, and calibration uncertainties are not included in the error bars; Column 8:Facilities; Column 9: References:This paper (L19); \citet{shao19} (S18); \cite{wang13} (W13); \cite{feruglio18} (F18), Riechers et al. in prep (Rpr)}

\label{table2}
\end{deluxetable*}

%\begin{rotate}
\begin{deluxetable*}{ccccc}
\tabletypesize{\scriptsize}
%\tablenum{3}
\tablecaption{Continuum Properties}
\tablecolumns{8}
\tablewidth{0pc}
%\tablecaption{Continuum detection with AMLA}
%\tablecaption{Continuum properties}
\tablehead{
\colhead{Frequency}  & \colhead{S$\nu$} &\colhead{Rms}&\colhead{Beam Size}  & \colhead{Source Size}\\ 
\colhead{(GHz)}  & \colhead{(mJy)} & \colhead{($\rm uJy\ beam^{-1}$)} &\colhead{(arcsec)}  & \colhead{(arcsec)} \\
\colhead{(1)}&\colhead{(2)}&\colhead{(3)}&\colhead{(4)}&\colhead{(5)} }
\startdata
80.6&0.22 $\pm$ 0.04& 16&1.68 $\times$ 1.37& (1.59 $\pm$ 0.52) $\times$ (0.21 $\pm$ 0.45)\\
96.0& 0.29 $\pm$ 0.03 &15& 1.42 $\times$ 1.19 & (0.83 $\pm$ 0.25) $\times$ (0.31 $\pm$ 0.39)\\
136.6 & 1.28 $\pm$ 0.03&15 &0.75 $\times$ 0.72 & (0.34 $\pm$ 0.04) $\times$ (0.22 $\pm$ 0.06) \\
141.1& 1.42 $\pm$ 0.03& 15& 0.80 $\times$ 0.65 & (0.27 $\pm$ 0.03) $\times$ (0.22 $\pm$ 0.06) \\
200.9& 3.88 $\pm$ 0.04 & 45&2.06 $\times$ 1.62 &-\\
215.9& 4.46 $\pm$ 0.05 &45 &1.92 $\times$ 1.54 &-
%289.2& 11.55 $\pm$ 0.08& 0.02 &0.48 $\times$ 0.39 & (0.27 $\pm$ 0.01) $\times$ (0.21 $\pm$ 0.01)&\\
%344.2&14.46 $\pm$ 0.12&0.04& 0.42 $\times$ 0.34 & (0.23 $\pm$ 0.01) $\times$ (0.19 $\pm$ 0.00) &\\
%\hline
%$f_{70\mu m}$ & 2.52\\
%$f_{160\mu m}$ & 1.24\\
%\hline
%$\rm {L_{\rm IR }}_{(8-1000\mu m)} $&&&&&$1.28_{-0.28}^{+0.33}$\\
%$\rm {L_{\rm FIR}}_{(42.5-122.5\mu m)} $&&&&&$0.99_{-0.21}^{+0.24}$
\enddata
\tablecomments{Column 1: Continuum frequency in observed frame; Column 2-3: Continuum flux density and rms; Column 4-5: Beam size and source size deconvolved from beam in FWHM. }
\label{table3}
\end{deluxetable*}

\begin{deluxetable*}{lcccc}
\tabletypesize{\scriptsize}
%\tablenum{4}
\tablecaption{MOLPOP-CEP Grid Parameter Ranges}
\tablecolumns{5}
\tablewidth{0pc}
%\tablecaption{Model parameter range}
\tablehead{\colhead{Input Parameters}& \colhead{Range} & \colhead{Grid Step} &{Grid Number}& \colhead{Unit}\\
\colhead{(1)}&\colhead{(2)}&\colhead{(3)}&\colhead{(4)}&\colhead{(5)} }
\startdata
Kinetic Temperature ($T_{\rm kin}$)&20 - 800& $\rm \Delta T_{kin} = 13$ &61&K\\
Volume Density ($ n(\rm H_{2})$)&$\rm 10^{3}$ - $\rm 10^{8}$&$\rm \Delta Log(n_{H2}/\rm cm^{-3}) = 0.25 $ &21&$\rm cm^{-3}$\\
Column Density ($N(\rm CO)$)&$\rm 10^{14}$ - $\rm 10^{21}$&$\rm \Delta Log(N_{CO}/\rm cm^{-2}) = 0.5$&15&$\rm cm^{-2}$
%Background Temperature ($\rm T_{bac}$) &19.12&-&-&K
%Line Width ($\rm \Delta V$)& 1&-&-&$\rm km\ s^{-1}$
\enddata
\tablecomments{Column 1: Input parameters to generate the grid; Column 2: Parameter ranges; Column 3: Steps of parameters in log space; Column 4: The resulting number of grid for a specific parameter; Column 5: Units}
\label{table4}
\end{deluxetable*}

\section{acknowledgement}
This work was supported by the National Science Foundation of China (11721303, 11991052) and the National Key R\&D Program of China (2016YFA0400702).
We acknowledge Andres Asensio Ramos for his substantial help on the code MOLPOP-CEP. We thank Chentao Yang, Roberto Decarli, Toshiki
Saito and Daizhong Liu for important discussions.
RW acknowledge supports from the Thousand Youth
Talents Program of China, the NSFC grants No. 11533001, and 11473004.
D.R. acknowledges support from the National Science Foundation under
grant numbers AST-1614213 and AST-1910107 and from the Alexander von
Humboldt Foundation through a Humboldt Research Fellowship for
Experienced Researchers.
YG's research is supported by National Key Basic Research and Development Program of China (grant No. 2017YFA0402704),
National Natural Science Foundation of China (grant Nos. 11861131007, 11420101002), and Chinese Academy of Sciences Key
Research Program of Frontier Sciences (grant No. QYZDJSSW-SLH008).
This paper is based on ALMA observations:ADS/JAO.ALMA 2015.1.01265.S. ALMA is a partnership of ESO (representing its member states), NSF (USA) and NINS (Japan), together with NRC (Canada), MOST and ASIAA (Taiwan), and KASI (Republic of Korea), in cooperation with the Republic of Chile. The Joint ALMA Observatory is operated by ESO, AUI/NRAO and NAOJ.
This paper also makes use of observations of  IRAM NOEMA Interferometer: Project number W18EE. IRAM is supported by INSU/CNRS (France), MPG (Germany) and IGN (Spain).

% tables in the paper


%\altaffiltext{1}{Visiting Astronomer, Cerro Tololo Inter-American Observatory.
%CTIO is operated by AURA, Inc.\ under contract to the National Science
%\altaffiltext{2}{Society of Fellows, Harvard University.}




\end{document}